\documentclass[11pt]{article}
\usepackage{axodraw}
\usepackage{epsfig}
\usepackage{amsfonts}
\usepackage{amsmath}
\usepackage{bbm}
\usepackage{cite}
\allowdisplaybreaks[1]

\input pix.sty

\newcommand{\Ao}{C}
\newcommand{\Bo}{D}
\newcommand{\Hc}{\mbox{H.c.}}

\newcommand{\nS}{n_\rmii{$S$}}
\newcommand{\nG}{n_\rmii{$G$}}
\newcommand{\mW}{m_\rmii{$W$}}
\newcommand{\mZ}{m_\rmii{$Z$}}

\newcommand{\mWt}{m_\rmii{$\widetilde W$}}
\newcommand{\mZt}{m_\rmii{$\widetilde Z$}}
\newcommand{\mQt}{m_\rmii{$\widetilde Q$}}

\renewcommand{\eq}{eq.~}
\renewcommand{\eqs}{eqs.~}

\renewcommand{\se}{sec.~}
\renewcommand{\ses}{secs.~}
\renewcommand{\fig}{fig.~}

\newcommand{\rmO}{{\mathcal{O}}}
\newcommand{\bmu}{\bar\mu}

\def\lsi{\raise0.3ex\hbox{$<$\kern-0.75em\raise-1.1ex\hbox{$\sim$}}}
\def\gsi{\raise0.3ex\hbox{$>$\kern-0.75em\raise-1.1ex\hbox{$\sim$}}}
\newcommand{\lsim}{\mathop{\lsi}}
\newcommand{\gsim}{\mathop{\gsi}}

\newcommand{\nF}{n_\rmii{F}}
\newcommand{\nB}{n_\rmii{B}}
 \renewcommand{\nF}[1]{n_\rmii{F{#1}}}
 \renewcommand{\nB}[1]{n_\rmii{B{#1}}}

\newcommand{\rmii}[1]{{\mbox{\tiny\rm{#1}}}}
\newcommand{\rmiii}[1]{{\mbox{\tiny{$\scriptstyle{\rm#1}$}}}}
\newcommand{\re}{\mathop{\mbox{Re}}}
\newcommand{\im}{\mathop{\mbox{Im}}}

\newcommand{\Tint}[1]{{\hbox{$\sum$}\!\!\!\!\!\!\!\int\,}_{\!\!\!\!\raise-0.9ex\hbox{$\scriptstyle{#1}$}}}
\newcommand{\Tinti}[1]{{{\Sigma}\!\!\!\!\raise0.3ex\hbox{$\int$}_\rmii{${#1}$}}}

\newcommand{\bi}{\begin{itemize}}
\newcommand{\ei}{\end{itemize}}
\newcommand{\hide}[1]{ }

\newcommand{\ff}{\rmi{\sl f\,}}

\newcommand{\deltabar}{\delta\!\!\!\raise0.7ex\hbox{--}\,}
\def\TAsc(#1,#2)(#3,#4,#5)%
{\SetWidth{2.0}\CArc(#1,#2)(#3,#4,#5)\SetWidth{1.0}}
\def\Lwidth{3}

\def\TAgl(#1,#2)(#3,#4,#5){\SetWidth{2.0}\PhotonArc(#1,#2)(#3,#4,#5){\Lwidth}%
{6.283 #3 mul 360 div #4 #5 sub #4 #5 sub mul sqrt mul Tdensity mul}%
\SetWidth{1.0}}
\def\TLgl(#1,#2)(#3,#4){\SetWidth{2.0}\Photon(#1,#2)(#3,#4){\Lwidth}
{#1 #3 sub #1 #3 sub mul #2 #4 sub #2 #4 sub mul add sqrt Tdensity mul}%
\SetWidth{1.0}}
\newcommand{\piC}[1]{\;\parbox[c]{120pt}{\begin{picture}(120,60)(0,0)
\SetWidth{1.0}\SetScale{1.2} #1 \end{picture}}\; }
\renewcommand{\picc}[1]{\;\parbox[c]{60pt}{\begin{picture}(60,30)(0,0)
\SetWidth{1.0}\SetScale{1.3} #1 \end{picture}}\; }
\def\Lwidth{1.3}

\def\NRa{\piC{%
 \SetWidth{1.0} 
 \Line(68,42)(100,21.5)%
 \Line(68,-2)(100,18.5)%
 \Photon(83,32)(94,46){1.5}{5}
 \CBox(98,18)(102,22){Black}{Yellow}
}}
\def\NRb{\piC{%
 \SetWidth{1.0} 
 \Line(68,42)(100,21.5)%
 \Line(68,-2)(100,18.5)%
 \Photon(83,32)(94,46){1.5}{5}
 \CBox(98,18)(102,22){Black}{Yellow}
 \COval(88.5,39)(3,3)(0){Black}{Gray}
 \SetWidth{0.4}
 \Line(94,46)(101,44)%
 \Line(94,46)(90,51)%
}}
\makeatletter \@addtoreset{equation}{section} \makeatother
\renewcommand{\theequation}{\arabic{section}.\arabic{equation}}
\makeatletter
\renewcommand\section{\@startsection {section}{1}{\z@}%
                                   {-5.5ex \@plus -1ex \@minus -.2ex}
                                   {2.3ex \@plus.2ex}%
                                   {\normalfont\large\bfseries}}
\renewcommand\subsection{\@startsection{subsection}{2}{\z@}%
                                     {-3.25ex\@plus -1ex \@minus -.2ex}%
                                     {1.5ex \@plus .2ex}%
                                     {\normalfont\normalsize\bfseries}}
\renewcommand\thesection {\@arabic\c@section}
\renewcommand\thesubsection   {\thesection.\@arabic\c@subsection}
\renewcommand{\@seccntformat}[1]{%
\csname the#1\endcsname.\hspace{1.0em}}
\makeatother

\begin{document}

\flushbottom

\begin{titlepage}

\begin{flushright}
July 2017
\vspace*{1cm}
\end{flushright} 
\begin{centering}

\vfill

{\Large{\bf
  Re-derived overclosure bound for the inert doublet model 
}} 

\vspace{0.8cm}

S.~Biondini and 
M.~Laine 

\vspace{0.8cm}

{\em
AEC, Institute for Theoretical Physics, 
University of Bern, \\ 
Sidlerstrasse 5, CH-3012 Bern, Switzerland\\} 

\vspace*{0.8cm}

\mbox{\bf Abstract}

\end{centering}

\vspace*{0.3cm}
 
\noindent
We apply a formalism accounting for thermal effects (such as modified
Sommerfeld effect; Salpeter correction; decohering scatterings; dissociation
of bound states), to one of the simplest WIMP-like dark matter models,
associated with an ``inert'' Higgs doublet. A broad temperature range 
$T \sim M/20 ... M/10^4$ is considered, stressing the importance and 
less-understood nature of late annihilation stages. Even though only 
weak interactions play a role, we find that resummed real and virtual 
corrections increase the tree-level overclosure bound by $1 ... 18\%$, 
depending on quartic couplings and mass splittings.

\vfill

 
\noindent

\vfill

\end{titlepage}

%
\section{Introduction}

Tight constraints from the LHC and from direct and 
indirect detection experiments have put many simple dark matter models
under tension in recent years. This calls 
for new ideas in model building, but perhaps 
also for new precision in the computations on which
a given dark matter scenario is based. Indeed, as the LHC 
pushes up the dark matter mass scale, it also increases the 
temperature at which dark matter density was fixed. Then, however, 
Standard Model weak interactions, which play a role in most 
dark matter computations, can be modified by thermal 
effects. If the freeze-out temperature is $T \gsim 160$~GeV,  
the Higgs mechanism ``melts away''~\cite{dono}, whereby weak interactions
display phenomena normally only associated with strong interactions. 

The purpose of this paper is to present a step-by-step 
implementation of a formalism which can 
account for relevant thermal effects~\cite{threshold},\footnote{%
 Other discussions of thermal effects relevant for heavy particles 
 can be found e.g.\ in refs.~\cite{dhr,therm1,therm2,thermal,therm3,therm4}. 
 }
and whose principal applicability has been tested against 
non-perturbative lattice simulations by using the annihilation of 
heavy quarks in QCD as an analogue 
for dark matter annihilation~\cite{4quark_lattice}. 
Among our goals are  
to check whether thermal modifications affect the
well-known Sommerfeld enhancement 
(cf.\ e.g.\ refs.~\cite{fadin,hisano,sfeldx,feng}), 
and how to include the classic  
Salpeter correction (cf.\ e.g.\ ref.~\cite{lsb}). 

The premise of the framework is 
to make use of a heavy-mass or ``non-relativistic'' 
expansion for the dark matter particles. 
Given that in the classic WIMP paradigm dark matter gradually 
freezes out at a temperature $T \sim M/20 ... M/10^4$, where $M$ is the
dark matter mass scale, there should be no doubt about the 
validity of this approximation. 

Within the non-relativistic regime, 
the framework accounts for a number
of thermal effects, such as that the vacuum masses of $W^\pm,Z^0$
are replaced by thermal Debye masses as the temperature
increases; that the weak mixing
angle evolves with the temperature; that weak interactions mediate
fast scatterings of the dark matter particles, 
transforming them into each other and thereby affecting the nature
of their annihilation process; that similar interactions also change the
effective mass of the dark matter particles through the Salpeter
correction; and that in some cases 
dark matter particles can form bound states. 
As far as the co-annihilation of non-degenerate dark
matter particles goes, the formalism can also be nicely
contrasted with the classic Boltzmann 
equation approach of ref.~\cite{old1}. 

To put the study in context, we remark that there has been 
recent interest in including next-to-leading order (NLO)
corrections into dark matter computations. Here we 
are more concerned with the fact that most computations
are formally incomplete even at leading order (LO), 
as far as near-threshold thermal effects go~\cite{threshold}. 
In principle, the inclusion of NLO corrections 
is also possible within the same formalism,
notably by adding operators suppressed by $\sim \nabla^2/M^2$ to 
\eq\nr{L} and NLO corrections to the coefficients given in  
\eqs\nr{c1}--\nr{c3}, however this is not pursued here. 

The model with which we choose to illustrate 
the formalism is a simple extension of the Standard Model through an 
additional ``inert'' Higgs doublet~\cite{id0,id1,id2}. Many dark
matter computations have been carried out for various parameter
corners of this model (cf.\ e.g.\ refs.~\cite{%
minimal,mhgt1,edsjo,mhgt2,agrawal,andreas,multiplet,
arina,dolle,ll1,schannel, 
pole,ai1,mk,arhrib,modak,queiroz,ibarra,banerjee,belyaev,nx1,nx2,nx3} 
and references therein; 
we particularly recommend ref.~\cite{multiplet} for a general overview), 
and our conclusions do not 
differ qualitatively from these, even though visible effects
from hitherto unconsidered processes can be observed. 

The plan of this paper is the following. After introducing the 
4-particle operators that mediate dark matter
decays in the heavy-mass limit (\se\ref{se:4part}), we recall how
they determine the thermal dark matter 
annihilation rate (\se\ref{se:rate}). 
Subsequently the key tools of the formalism, namely time-dependent 
medium-modified Schr\"odinger equations governing the ``slow'' 
dynamics within the dark sector, are elaborated upon (\se\ref{se:S}). 
After presenting numerical solutions and the overclosure
bound (\se\ref{se:numerics}), 
we turn to conclusions and an outlook (\se\ref{se:concl}). 

%
\section{4-particle operators}
\la{se:4part}

In the inert doublet model (IDM), the Standard Model is supplemented
by an additional Higgs doublet, $\chi$, which does not couple to fermions
because of an unbroken discrete Z(2) symmetry. Denoting by $\phi$ the 
Standard Model Higgs doublet and by $D^{ }_\mu$ the corresponding
covariant derivative, the Standard Model Lagrangian 
is modified by the additional terms
\ba
 \mathcal{L}_\chi & = &  
 (D^\mu \chi)^\dagger (D_\mu\chi) - M^2 \chi^\dagger \chi \nn  
 & - &  \biggl\{  
   \lambda^{ }_2\, ( \chi^\dagger \chi^{ } )^2   
 + \lambda^{ }_3\, \phi^\dagger \phi^{ }\, \chi^\dagger \chi^{ }
 + \lambda^{ }_4\, \phi^\dagger \chi^{ }\, \chi^\dagger \phi^{ }
 + \biggl[ \frac{\lambda^{ }_5}{2}\, (\phi^\dagger \chi^{ })^2 + \Hc\biggr]
 \biggr\} 
 \;. \la{V} 
\ea
The notation $\lambda^{ }_1$ is reserved for the Standard Model
Higgs self-coupling, 
$
 \delta \mathcal{L}^{ }_\rmii{SM} = - \lambda^{ }_1 (\phi^\dagger \phi)^2
$.

If the mass scale $M$ is much larger than the electroweak scale, 
$M \gg \mW$, the $\chi$ particles annihilate efficiently into the 
Standard Model ones. The annihilations that we are interested
in happen in the temperature range $T \sim M/20 ... M/10^4$, 
in which the average velocity is $v \sim \sqrt{T/M} \ll 1$. Therefore
the annihilating particles are non-relativistic. 
Non-relativistic annihilations can be described by
4-particle operators, arranged as an expansion in $1/M^2$~\cite{bodwin}.  
If we write non-relativistic on-shell fields in terms of 
annihilation and creation operators as 
\be
 \chi = \frac{1}{\sqrt{2M}}
 \Bigl( \Ao e^{- i M t} + \Bo^\dagger e^{i M t}\Bigr) 
 \;, \quad
 \chi^\dagger = \frac{1}{\sqrt{2M}}
 \Bigl( \Bo e^{- i M t} + \Ao^\dagger e^{i M t}\Bigr) 
 \;, 
\ee
then at leading order in $1/M^2$ there are four ``absorptive'' 
operators that play a role:\footnote{%
  As is characteristic of an effective theory approach, there are
  in principle infinitely many higher-dimensional operators, suppressed
  by increasing powers of $1/M^2$. The four operators here are the
  only ones at order $1/M^2$.
  The coefficients of these operators contain both a real part and 
  an imaginary (i.e.\ absorptive) part~\cite{bodwin}.
  Only the imaginary parts are relevant for us~\cite{threshold}:
  in accordance with
  the optical theorem, they represent matrix elements squared of
  real processes in which the heavy particles annihilate into Standard Model
  ones. The annihilations are two-particle annihilations; therefore
  the matrix elements squared contain four field operators,
  two annihilation operators for a process, and two creation
  operators for its conjugate. In eqs.~\nr{c1}--\nr{c3} the coefficients
  of these operators are given at leading order, corresponding to a
  tree-level annihilation cross section. One strength of the
  effective theory approach is that if needed, it would be fairly
  straightforward to compute NLO corrections to the coefficients.
  Even more importantly, soft thermal corrections to the annihilation
  processes (cf.\ \fig\ref{fig:diagram} for an illustration) 
  can be included beyond a quasi-particle approximation,
  and up to the non-perturbative level in the case
  of strong interactions~\cite{4quark_lattice}.
 } 
\be
 \delta \mathcal{L}^{ }_\rmi{abs} = i \, 
 \bigl( 
  c^{ }_1 \, \underbrace{ 
    \Ao^\dagger_p \Bo^\dagger_p \Bo^{ }_q \Ao^{ }_q
    }_{\equiv\; O^{ }_1}
 \,+\, 
  c^{ }_2 \, \underbrace{ 
    \Ao^\dagger_p T^a_{pq} \Bo^\dagger_q\, \Bo^{ }_r T^a_{rs} \Ao^{ }_s
    }_{\equiv\; O^{ }_2}
 \,+\, 
  c^{ }_3 \, \underbrace{ 
    \Bo^\dagger_p \Bo^\dagger_q \Bo^{ }_p \Bo^{ }_q
    }_{\equiv\; O^{ }_3}
 \,+\, 
  c^{ }_4 \, \underbrace{ 
    \Ao^\dagger_p \Ao^\dagger_q \Ao^{ }_p \Ao^{ }_q
    }_{\equiv\; O^{ }_4}
 \; \bigr)
 \;. \la{L}
\ee 
Here sums over the isospin components
$p,q,r,s \in \{1,2\}$ are implied, and 
$T^a \equiv \sigma^a/2$, where $\sigma^a$ are the Pauli matrices.

We have computed the coefficients $c^{ }_1,...,c^{ }_4$ 
in \eq\nr{L} in general 
$R^{ }_{\xi}$ gauges at leading non-trivial order, 
verifying their gauge independence for
$\xi < M^2 / \mZ^2$:\footnote{%
 For
 $\xi \gg M^2 / \mZ^2 \gg 1$ the results change qualitatively 
 and therefore  unitary gauge is not viable.} 
\ba
 c^{ }_1 & = & \frac{
 g_1^4 + 3 g_2^4 + 8 \lambda_3^2 + 8 \lambda^{ }_3 \lambda^{ }_4
  + 2 \lambda_4^2 
 }{256\pi M^2}
 \;, \la{c1} \\ 
 c^{ }_2 & = & \frac{ 
 g_1^2 g_2^2 + \lambda_4^2 
 }{32\pi M^2}
 \;, \la{c2} \\  
 c^{ }_3 & = & c^{ }_4 \; = \; \frac{
 \lambda_5^2
 }{128\pi M^2}
 \;. \la{c3}
\ea
Here $g^{ }_1$ and $g^{ }_2$ are the U$^{ }_\rmii{Y}$(1)
and SU$^{ }_\rmii{L}$(2) gauge couplings, respectively. 
The couplings should be evaluated 
at a renormalization scale $\sim 2M$.
The same values of the coefficients can be extracted
from ref.~\cite{multiplet}. 

If $\lambda^{ }_4 \neq 0$ or $\lambda^{ }_5 \neq 0$, or if Standard Model
radiative corrections are considered, different components of $\chi$ are
non-degenerate in mass. In this case the doublets $\Ao$ and $\Bo$ can be 
written as 
\be
 \Ao = \left( 
     \begin{array}{c} H^{ }_+ \\[1mm] 
     {\displaystyle \frac{H^{ }_0-iH^{ }_{\bar{0}}}{\sqrt{2}} }
     \end{array} 
     \right)
 \;, \quad 
 \Bo = \left( 
     \begin{array}{c} H^{ }_- \\[1mm] 
     {\displaystyle \frac{H^{ }_0+iH^{ }_{\bar{0}}}{\sqrt{2}} }
     \end{array} 
     \right)
 \;. \la{def_H}
\ee
The operators in \eq\nr{L} split into a $10 \times 10$ matrix 
in the field space of \eq\nr{def_H}, which is given (with a slightly
different notation) in \eqs\nr{W1}--\nr{W4} below.

%
\section{Rate equations and effective cross sections}
\la{se:rate}

As discussed in ref.~\cite{old1}, 
the only physically reasonable ``slow variable'' 
of the problem at hand is the {\em total} number density of dark matter 
particles, $n \equiv \sum_{i=\pm,0,\bar{0}} n^{ }_i$. Within a Boltzmann 
approach, ref.~\cite{old1} established that $n$ evolves according
to the Lee-Weinberg equation~\cite{clas1,clas2}, 
\be
 \dot{n} \; = \; -\langle \sigma^{ }_\rmi{eff}\, v \rangle \, 
 \bigl( n^2 - n_\rmi{eq}^2 \bigr) 
 \;, \la{boltzmann}
\ee
where $\dot{n}$ is the covariant time derivative in an expanding 
background, and 
\be
 \langle \sigma^{ }_\rmi{eff}\, v \rangle \; = \;
 \sum_{i,j} \frac{
 \langle \sigma^{ }_{ij} v^{ }_{ij} \rangle
  n_i^\rmi{eq} n_j^\rmi{eq} }
 {n_\rmi{eq}^2} \la{sigma_eff}
\ee
is an effective cross section for $2\to 2$ annihilations from the dark sector. 
In our case the total equilibrium number density reads, at tree-level,  
\be
 n^{ }_\rmi{eq} 
 \; \approx \; \int_{\vec{k}}
 \Bigl( e^{-E^{ }_+/T} + e^{-E^{ }_-/T} + e^{-E^{ }_0/T}
 + e^{-E^{ }_{\bar{0}}/T} \Bigr) 
 \; = \; 
 \sum_{i=\pm,0,\bar{0}}  
 \frac{T M_{H_i}^2}{2\pi^2} K^{ }_2\biggl( \frac{M^{ }_{H_i}}{T} \biggr)
 \;, \la{neq}
\ee
where 
$ 
 \int_{\vec{k}} \equiv \int \frac{{\rm d}^3\vec{k}}{(2\pi)^3}
$,
$
 E^{ }_i \equiv \sqrt{k^2 + M_{H_i}^2}
$
with $k \equiv |\vec{k}|$, 
and $K^{ }_2$ is a modified Bessel function.

We note in passing that  
radiative corrections to \eq\nr{neq} can be determined 
as explained in ref.~\cite{threshold}. The most important 
is the so-called Salpeter correction, which modifies the
{\em rest mass} of a non-relativistic particle by an 
amount $\Delta M^{ }_\rmii{$T$} \sim -\alpha^{3/2} T < 0 $, where $\alpha$
is a weak fine-structure constant (cf.\ e.g.\ ref.~\cite{therm2}). 
This is specified in more detail in \se\ref{se:numerics}
(cf.\ \eq\nr{neq2}). 

In contrast to \eq\nr{boltzmann}, 
the formalism of ref.~\cite{threshold}
takes as a starting point an equation based on general linear 
response theory, having thus the form~\cite{chemical} 
\be
 \dot{n} \; = \; -\Gamma^{ }_\rmi{chem} \bigl( n - n^{ }_\rmi{eq} \bigr) 
 + \rmO \bigl( n - n^{ }_\rmi{eq} \bigr)^2
 \;, \la{response}
\ee
where $\Gamma^{ }_\rmi{chem}$ can be called the chemical equilibration rate. 
In the remainder of this paper, we wish to make close contact 
with standard literature, and therefore prefer to use the form 
of \eq\nr{boltzmann}. Linearizing \eq\nr{boltzmann} in deviations from
equilibrium leads us to identify 
\be
 \langle \sigma^{ }_\rmi{eff}\, v \rangle \equiv 
 \frac{\Gamma^{ }_\rmi{chem}}{2 n^{ }_\rmi{eq}}
 \;. \la{sigma_2}
\ee
In the absence of a first-principles argument beyond the linear response
level, we rely on the form of \eq\nr{boltzmann} on how first and higher order 
deviations are related to each other. 

The strength of the linear response approach is that it permits to 
relate the equilibration rate $\Gamma^{ }_\rmi{chem}$ to a correlator 
evaluated in equilibrium, without assuming weak interactions or 
the validity of a quasi-particle description necessary for 
a Boltzmann treatment~\cite{chemical}. 
Specifically, when the reactions responsible
for equilibration are described by operators of the type in 
\eq\nr{L}, $\Gamma^{ }_\rmi{chem}$ is 
to first order proportional to the thermal 
expectation value of $\delta\mathcal{L}^{ }_\rmi{abs}$~\cite{4quark_lattice}. 
Inserting the proportionality coefficient and 
expressing the result through \eq\nr{sigma_2}, we obtain 
\be
 \langle \sigma^{ }_\rmi{eff} \, v \rangle = 
 \frac{4}{n_\rmi{eq}^2}
   \sum_{i=1}^{4} c^{ }_i \gamma^{ }_i
 \;, \quad
 \gamma^{ }_i \; \equiv \; 
 \bigl\langle
   O^{ }_i  
 \bigr\rangle^{ }_{ }
 \;. \la{sigma}
\ee
Because the annihilation operators are positioned to the right
in \eq\nr{L}, the vacuum state does not contribute to the 
expectation value in \eq\nr{sigma}. Therefore $\gamma^{ }_i$
is exponentially suppressed by $\sim e^{-2M/T}$, with the Boltzmann
factor cancelling against that from $n_\rmi{eq}^2$.

Eq.~\nr{sigma} represents 
a generalization of \eq\nr{sigma_eff}. The matrix
structure of $\sigma^{ }_{ij}$ corresponds to matrix-like
Schr\"odinger equations satisfied by the wave functions of the 
annihilating pair (cf.\ table~\ref{table:potentials}), 
and the weights $n^\rmi{eq}_i$ in \eq\nr{sigma_eff}
correspond to threshold
locations in the Laplace transform in \eq\nr{Laplace2}. 
At the same time  
\eq\nr{sigma} goes beyond \eq\nr{sigma_eff} in several respects, 
for instance by permitting
for a systematic inclusion of virtual thermal effects 
in the computation of individual cross sections, and also of
real thermal scatterings of the 
dark matter particles off Standard Model particles, 
as discussed in more detail
in \ses\ref{ss:general} and \ref{ss:applicable}.

%
\section{Schr\"odinger description}
\la{se:S}

%
\subsection{General goal and physical interpretation}
\la{ss:general}

In the notation of ref.~\cite{old1}, 
the cross sections in \eq\nr{sigma_eff} describe the processes
\be
 \chi^{ }_i\chi^{ }_j \leftrightarrow X X'
 \;, \la{reactions0}
\ee
where $X,X'$ are Standard Model particles. 
These are ``slow'' processes: the likelihood that a dark matter
particle finds a partner with which to annihilate is suppressed
by a Boltzmann factor, so that the rate is  
$\Gamma \sim \frac{\alpha^2}{M^2} \int_{\vec{k}} e^{-E/T}$. 
However, the 
$\chi^{ }_i$-particles also experience ``fast'' reactions which
have no Boltzmann suppression associated with them. 
These are of the type given in \eqs(6b) and (6c) of ref.~\cite{old1}: 
\be
 \chi^{ }_i X \leftrightarrow \chi^{ }_j X' \;, \quad 
 \chi^{ }_i \leftrightarrow \chi^{ }_j X X' \;. 
 \la{reactions} 
\ee 
These reactions 
keep the dark matter particles in kinetic equilibrium, and also
change them into each other,  guaranteeing chemical 
equilibrium within the dark sector, with each species contributing
with its proper number density $n_\rmi{eq}^i$ into \eq\nr{sigma_eff}. 
If there are bound states in the dark sector, further ``fast'' processes
can be added, notably
\be
 (\chi^{ }_i \chi^{ }_k)^{ }_\rmi{open} X 
 \leftrightarrow 
 ( \chi^{ }_j \chi^{ }_l)^{ }_\rmi{bound} X' \;, \quad 
 ( \chi^{ }_i \chi^{ }_k)^{ }_\rmi{open} 
 \leftrightarrow 
 ( \chi^{ }_j \chi^{ }_l)^{ }_\rmi{bound} X X'  
 \;, \la{reactions2}
\ee
where we assume that the binding energy is small, 
$\Delta E \sim \alpha^2 M \lsim \pi T$. Of course the same reactions are
also present without bound states, 
\be
 (\chi^{ }_i \chi^{ }_k)^{ }_\rmi{open} X 
 \leftrightarrow 
 ( \chi^{ }_j \chi^{ }_l)^{ }_\rmi{open} X' \;, \quad 
 ( \chi^{ }_i \chi^{ }_k)^{ }_\rmi{open} 
 \leftrightarrow 
 ( \chi^{ }_j \chi^{ }_l)^{ }_\rmi{open} X X'  
 \;, \la{reactionsX}
\ee
and can change the annihilating pair into a different gauge or spin state.
In addition, processes with 
virtual $X$ exchange are important, 
 \be
 \chi^{ }_i\chi^{ }_j\,(\mbox{virtual~}X) \leftrightarrow X' X''
 \;, \la{reactions3}
\ee
leading e.g.\ to the Sommerfeld effect.

The description based on \eq\nr{sigma} goes beyond \eq\nr{sigma_eff} 
in that the indirect effect of the reactions 
in \eqs\nr{reactions}--\nr{reactions3} can be included in a more
``differential'' form.  
Specifically, the fast reactions in \eq\nr{reactions} give thermal
masses to the dark matter particles, which change the kinematics 
of the reactions in \eq\nr{reactions0}, leading e.g.\ to the 
Salpeter correction whereby the location of 
the 2-particle threshold gets modified. 
The fast reactions also induce thermal 
interaction rates, which decohere quantum-mechanical phases and
thereby affect cross sections. Likewise the Sommerfeld effect and 
the possible emergence of bound states are included, through the 
solution of dynamical (time-dependent) Schr\"odinger equations.
Thereby there is no need to assume the validity 
of a quasi-particle picture in the dark sector. 

%
\subsection{On the applicability of the Schr\"odinger description}
\la{ss:applicable}

Despite its strengths, 
an effective Schr\"odinger description as outlined 
in \se\ref{ss:general} is only
valid in a certain parametric regime. Indeed its
justification requires an analysis of 
the different energy and momentum scales contributing to the problem. 
For near-threshold problems at finite temperature, several different
scales play a role. 
A thermally modified Schr\"odinger approach
in the form implemented below can be used for addressing
energy scales $\Delta E \sim \alpha^2 M$ provided that 
(cf.\ e.g.\ refs.~\cite{peskin,soto,jacopo,review})
\be
 \alpha^2 M \;\ll\; g T \,,\; \alpha M \,,\; \pi T \;\ll\; M
 \;, \la{scales}
\ee
where $\alpha \sim g^2/(4\pi)$. In this situation the scale $gT$,
which is the Debye scale representing typical energies/momenta
of soft Standard Model excitations, can be integrated out, so that no 
Standard Model fields appear in the description of the ``slow'' dynamics. 

An example of an excitation associated with the 
scale $gT$ is an electric dipole $\sim \vec{r}\cdot g\vec{E}$. 
As discussed in ref.~\cite{jacopo}, such 
dipoles cause transitions between pairs in different gauge representations, 
as appear in the operators of \eq\nr{L}. Specifically, integrating
out the $\vec{E}$ fields and the pairs in repulsive channels
generates a thermal interaction rate affecting the dynamics of  
the pair in an attractive channel~\cite{jacopo}. 

Now, the interaction rate in the attractive channel is 
a {\em slow} rate: the annihilating pair is in a gauge-singlet
state and only a dipole contribution is left over, 
$\Gamma \sim \alpha^2 T^3 r^2$. Therefore, $\Gamma$ can 
be part of an effective slow quantum-mechanical description.  

In contrast, the generic interaction rates in the Standard Model, 
and in particular the interaction rates of the heavy $\chi$ 
pairs in gauge non-singlet channels,  
are of order $\alpha T$. This is a {\em fast} rate, rapidly decohering 
the phase of the wave function and justifying a classical Boltzmann 
description. At the same time, it is not clear whether such a rate can be 
consistently included in a Schr\"odinger equation: if $T \gsim \alpha M$, 
$\Gamma\sim\alpha T$ modifies the spectral function in the energy range 
$\Delta E \sim \alpha^2 M$ by an effect of $\rmO(1)$, yielding 
a substantial below-threshold tail akin to 
that appearing below the top-antitop 
threshold in vacuum~\cite{khoze}.

We have adopted a procedure here in which 
the contributions of the repulsive channels are estimated in two 
ways: either including the below-threshold tail, or omitting it.
The difference of the results is used for estimating 
the theoretical uncertainties of our computation from 
thermal effects which are formally of NLO magnitude. 

\begin{figure}[t]

\hspace*{-0.1cm}
\centerline{%
 \epsfxsize=8.0cm\epsfbox{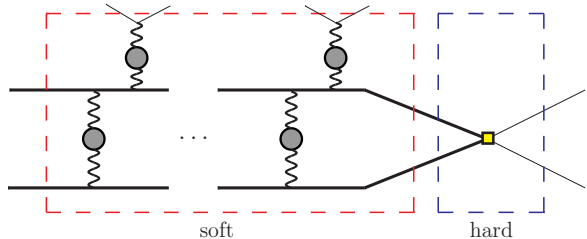}%
}

\caption[a]{\small
 An illustration of the repeated
 interactions of the dark matter particles with
 plasma constituents and each other, 
 before the annihilation into Standard Model
 particles takes place.
 Thick lines stand for dark matter particles, thin lines for
 Standard Model particles, and wiggly lines for gauge bosons. The blobs
 indicate that because of infrared sensitivity Hard Thermal Loop  
 resummed propagators
 need to be used for gauge bosons.
 The red (blue) dashed box encompasses 
 the soft (hard) interactions. 
 The soft interactions comprise both virtual and real corrections, 
 and the dots stand for
 iterations resummed through the Schr\"odinger
 description. 
 The hard process, with a large energy release of order $M$, 
 converts dark matter particles into Standard Model ones.
}

\la{fig:diagram}
\end{figure}

Having introduced the four-particle operators (cf.\ \eq\nr{L}) 
and the Schr\"odinger approach, we can briefly comment on the
different stages of the annihilation process. 
According to the scale hierarchy in \eq\nr{scales}, 
there are two well-separated classes of processes:
those occurring at the hard scale, $M$, and those typical of the soft
scales, either thermal or non-relativistic  
(cf.\ \fig\ref{fig:diagram}). 
The latter account for several
interactions with particles from the heat bath which are resummed by
a thermally modified Schr\"odinger equation.
In the end the dark matter particles annihilate into Standard Model ones.
This happens at a typical distance scale of order $1/M$ which is not
resolved by the larger medium length scales. 
Hence an effective point-like interaction is 
responsible for the hard process.
Such a factorization manifests itself 
in the effective cross section, 
\eq\nr{sigma_full} below, 
where the hard coefficients from \eqs\nr{c1}--\nr{c3} multiply 
thermal expectation values capturing the soft physics. 

%
\subsection{Degenerate limit}
\la{ss:degenerate}

We start by considering the degenerate limit, 
i.e.\ $M \equiv M^{ }_0 = M^{ }_{\bar{0}} = M^{ }_{\pm}$.
Each of the expectation values in \eq\nr{sigma}
can be expressed as a Laplace transform of a 
spectral function, denoted by $\rho^{ }_i$ 
(cf.\ \eqs\nr{Pismall}--\nr{rel2} below). 
Under the assumptions discussed in \se\ref{ss:applicable} 
and going over to non-relativistic
center-of-mass coordinates, 
the spectral function
is in turn an imaginary part of a Coulomb Green's function~\cite{threshold}: 
\ba
 \biggl[ 
   -\frac{\nabla_r^2}{M} + \mathcal{V}^{ }_i(r) - E'
 \biggr] G^{ }_i(E';\vec{r},\vec{r'}) & = & 
 N^{ }_i\, \delta^{(3)}(\vec{r}-\vec{r'})
 \quad \mbox{(no sum over $i$)}\;, \la{Seq1} \\ 
 \lim_{\vec{r,r'}\to \vec{0}} \im  G^{ }_i(E';\vec{r},\vec{r'})
 & = & \rho^{ }_i(E') \la{get_rho1}
 \;, 
\ea
where $N^{ }_i$ is a normalization factor giving the number of 
contractions related to $O^{ }_i$:
\be
 N^{ }_1 = 2 \;, \quad N^{ }_2 = \fr32 \;, \quad N^{ }_3 = N^{ }_4 = 6
 \;. \la{Ns}
\ee
In center-of-mass coordinates the Laplace transform reads  
\ba
 \gamma^{ }_i & \approx &
 \int_{\vec{k}} e^{-\frac{2M}{T} - \frac{k^2}{4 M T}}
 \int_{-\Lambda}^{\infty} \! \frac{{\rm d} E'}{\pi} \, e^{-E'/T}
 \; \rho^{ }_i(E') 
 \nn
 & = & 
 \Bigl( \frac{MT}{\pi} \Bigr)^{3/2} e^{-2M/T}
 \int_{- \Lambda}^{\infty} 
 \! \frac{{\rm d}E'}{\pi} \, e^{-E'/T}
 \; \rho^{ }_i(E')
 \;, \la{Laplace1}
\ea 
where $M \gg \Lambda \gg \alpha^2 M$ is a cutoff restricting the average
to the non-relativistic regime. According to \eq\nr{sigma}, the physical
result is $\sum_{i=1}^4 c^{ }_i \gamma^{ }_i$, with $c^{ }_i$
given in \eqs\nr{c1}--\nr{c3}. 

In the free limit, $\mathcal{V}^{ }_i\to 0$, 
the spectral function from \eqs\nr{Seq1} and \nr{get_rho1} reads 
$
 \rho^{(0)}_i(E') =  N^{ }_i M^{\fr32}\theta(E') \sqrt{E'}/(4\pi)
$.
Carrying out the Laplace transform in \eq\nr{Laplace1}, inserting 
$
 n^{(0)}_\rmi{eq} = 4 
 \bigl( \frac{MT}{2\pi} \bigr)^{\fr32} e^{-M/T}
$
from \eq\nr{neq}, and plugging into \eq\nr{sigma}, we obtain
the value of $ \langle \sigma^{ }_\rmi{eff}\, v \rangle $ 
for a degenerate system and
to leading order in $1/M^2$ and $\alpha$:
\be
 \langle \sigma^{ }_\rmi{eff}\, v \rangle^{(0)}_{ } = 
 \frac{c^{ }_1}{2} + \frac{3 c^{ }_2}{8} + \frac{3(c^{ }_3 + c^{ }_4)}{2}
 \;. \la{sigma_free}
\ee

In order to go beyond \eq\nr{sigma_free}, we include the potentials
$\mathcal{V}^{ }_i$ for the various channels in \eq\nr{Seq1}. 
It is helpful to introduce the notation 
\ba
 \mathcal{V}^{ }_\rmii{${W}{W}$}(r) & \equiv & 
 \frac{g_2^2}{4}
 \int_{\vec{k}} e^{i \vec{k}\cdot\vec{r}}
 \, i \langle W^{+}_0 {W}^{-}_0 \rangle^{ }_\rmii{T}(0,{k})
 \;, \la{V_WW} \\ 
 \mathcal{V}^{ }_\rmii{${A}{A}$}(r) & \equiv & 
 \frac{g_2^2}{4}
 \int_{\vec{k}} e^{i \vec{k}\cdot\vec{r}}
 \, i \langle A^{3}_0 {A}^{3}_0 \rangle^{ }_\rmii{T}(0,{k})
 \;,  \\
 \mathcal{V}^{ }_\rmii{${B}{B}$}(r) & \equiv & 
 \frac{g_1^2}{4}
 \int_{\vec{k}} e^{i \vec{k}\cdot\vec{r}}
 \, i \langle B^{ }_0 B^{ }_0 \rangle^{ }_\rmii{T}(0,{k})
 \;, \la{pot_deg}
\ea
where $\langle ... \rangle^{ }_\rmii{T}$ denotes a time-ordered propagator
and the gauge potentials have been expressed with the sign conventions of
the imaginary-time formalism. For instance 
(cf.\ appendix~A of ref.~\cite{threshold} for a derivation),\footnote{%
  For hard momenta $k \gg \mWt^{ }$ only the massless 
  part $1/k^2$ is important. The full form is needed for correctly 
  estimating the contribution of soft near-threshold momenta to the
  annihilation cross section. The soft momenta become increasingly
  important as the temperature decreases. 
 } 
\ba
 i 
 \bigl\langle W^{+}_0 W^{-}_0 \bigr\rangle^{ }_\rmii{T}(0,k)
 & = &   
 \frac{1}{k^2 + \mWt^2}
 \; - \; \frac{i\pi T}{k}
 \frac{m_\rmii{E2}^2}
        {(k^2 + \mWt^2)^2}
 \;, \hspace*{6mm} \la{W0_prop}
\ea
where 
$\mW = g^{ }_2 v/2$ is the $W^\pm$ mass, 
$v$ is the temperature-dependent Higgs expectation value,\footnote{%
  Even though carrying the same symbol,
  $v$ should not be confused with the non-relativistic velocity
  appearing e.g.\ in \eq\nr{boltzmann}.
 } 
and $\mWt^2 \equiv \mW^2 + m_\rmii{E2}^2$, 
where $m_\rmii{E2}^2$ is a Debye mass~\cite{meg}
(for future reference we also define $m_\rmii{E1}^2$ here): 
\be
 m^{2}_\rmii{E1} \; \equiv \; 
 \Bigl( \fr{\nS}6 + \frac{5\nG}{9} \Bigr) g_1^2 T^2 
 \;, \quad
 m^{2}_\rmii{E2} \; \equiv \; 
 \Bigl( \fr23 + \fr{\nS}6 + \frac{\nG}{3} \Bigr) g_2^2 T^2
 \;, \quad
 \nS \equiv 1
 \;, \quad
 \nG \equiv 3
 \;. \la{Debye}
\ee
For the neutral gauge field components ($B^{ }_0,A^{3}_0$)
the propagator is a matrix, whose form can 
be found in \eqs(A.22) and (A.23) of ref.~\cite{threshold}. 

With the notation introduced, the potentials appearing in 
\eq\nr{Seq1} read
\ba
 \mathcal{V}^{ }_1(r) 
 & = & 
    2 \mathcal{V}^{ }_\rmii{${W}{W}$}(0)
   + \mathcal{V}^{ }_\rmii{${A}{A}$}(0)  
   + \mathcal{V}^{ }_\rmii{${B}{B}$}(0) 
   - 2 \mathcal{V}^{ }_\rmii{${W}{W}$}(r)
   - \mathcal{V}^{ }_\rmii{${A}{A}$}(r)  
   - \mathcal{V}^{ }_\rmii{${B}{B}$}(r) 
 \;, 
 \la{pot1} \\[3mm]
 \mathcal{V}^{ }_2(r) 
 & = & 
      2 \mathcal{V}^{ }_\rmii{${W}{W}$}(0)
   + \mathcal{V}^{ }_\rmii{${A}{A}$}(0)  
   + \mathcal{V}^{ }_\rmii{${B}{B}$}(0) 
   + \frac{2 \mathcal{V}^{ }_\rmii{${W}{W}$}(r)
          +  \mathcal{V}^{ }_\rmii{${A}{A}$}(r)}{3}  
   - \mathcal{V}^{ }_\rmii{${B}{B}$}(r) 
 \;, 
 \la{pot2} \\[3mm] 
 \mathcal{V}^{ }_{3,4}(r)
 & = & 
     2 \mathcal{V}^{ }_\rmii{${W}{W}$}(0)
   + \mathcal{V}^{ }_\rmii{${A}{A}$}(0)  
   + \mathcal{V}^{ }_\rmii{${B}{B}$}(0) 
   + \frac{2 \mathcal{V}^{ }_\rmii{${W}{W}$}(r)
   + \mathcal{V}^{ }_\rmii{${A}{A}$}(r)}{3}
   + \mathcal{V}^{ }_\rmii{${B}{B}$}(r) 
 \;. \la{pot3} \hspace*{10mm}
\ea
The $r$-independent parts, denoted somewhat formally 
with the argument $r=0$, correspond to self-energy contributions; 
the $r$-dependent parts to exchange contributions.\footnote{%
 The $r$-dependent parts vanish at $r\to\infty$, so that 
 $
   \lim_{r\to\infty} \mathcal{V}^{ }_i(r) = 
     2 \mathcal{V}^{ }_\rmii{${W}{W}$}(0)
   + \mathcal{V}^{ }_\rmii{${A}{A}$}(0)  
   + \mathcal{V}^{ }_\rmii{${B}{B}$}(0) 
 $.
 } 
The $r$-independent parts 
are linearly divergent, and the corresponding
vacuum counterterms are defined
such that $\lim_{r\to \infty}\mathcal{V}^{ }_i(r) = 0$
at $T = 0$. Explicit expressions are given in appendix~A.
At $T > 0$, 
$\lim_{r\to \infty}\re[\mathcal{V}^{ }_i(r)] \neq 0$
amounts to the Salpeter correction. 
As elaborated upon in \se\ref{ss:applicable} and as can be deduced
from \eq\nr{pot1}, 
in $\mathcal{V}^{ }_{1}$ the thermal widths cancel
to leading order in $r \sim 1/(Mv)$, whereas in 
$\mathcal{V}^{ }_{2,3,4}$ they represent
fast reaction rates $\sim \alpha T$.

%
\subsection{Non-degenerate situation}
\la{ss:nondeg}

If $\lambda^{ }_4 \neq 0$ or $\lambda^{ }_5 \neq 0$ and $v > 0$, 
\eq\nr{V} implies that different components of the inert doublet
$\chi$ have different masses. A mass splitting is also induced
by Standard Model radiative corrections~\cite{minimal}. 
In this situation the potentials
of \eqs\nr{pot1}--\nr{pot3} get replaced by matrix potentials which
act in the space of the field components $H^{ }_{\pm}$, 
$H^{ }_0$, $H^{ }_{\bar{0}}$ defined in \eq\nr{def_H}. 
Modifying the notation slightly from \eq\nr{L}, 
we denote the mass of the neutral component $H^{ }_0$ by $M$, 
and the additional rest mass of the pair $H^{ }_i H^{ }_j$
by $\Delta M^{ }_{ij}$. The kinetic masses appearing
in the Schr\"odinger equations also depend on the pair 
in question, however for small but non-zero 
$\Delta M^{ }_{ij} \gsim \alpha^2 M$ this can be 
considered to be a higher-order effect, and will 
be omitted in the following (its inclusion
is trivial, by replacing the kinetic term in \eq\nr{Seq} by 
a diagonal matrix containing the reduced masses). 

Even though \eq\nr{sigma} contains expectation values of the type
\ba
 \gamma^{ }_i & = & 
 \int_{-\infty}^{\infty} \! \frac{{\rm d}\omega}{2\pi}\, \int_{\vec{k}}
 \Pi^{<}_i(\omega,\vec{k}) \;, \\ 
 \Pi^{<}_1(\omega,\vec{k})  & \equiv & 
 \int_{-\infty}^{\infty} \! {\rm d}t \, e^{i\omega t} \! \int_{\vec{r}}
 e^{-i \vec{k}\cdot\vec{r}} 
 \bigl\langle\,
          (\Ao^\dagger_p \Bo^\dagger_p)(0,\vec{0}) \,
          (\Bo^{ }_q \Ao^{ }_q)(t,\vec{r})  
 \,\bigr\rangle
 \;, \la{Pismall}
\ea
for the Schr\"odinger equation it is convenient to consider the opposite
time ordering~\cite{peskin}, 
\be
 \Pi^{>}_1(\omega,\vec{k}) \;\equiv\;
 \int_{-\infty}^{\infty} \! {\rm d}t \, e^{i\omega t} \! \int_{\vec{r}}
 e^{-i \vec{k}\cdot\vec{r}} 
 \bigl\langle\,
          (\Bo^{ }_q \Ao^{ }_q)(t,\vec{r})\,  
          (\Ao^\dagger_p \Bo^\dagger_p)(0,\vec{0}) 
 \,\bigr\rangle
 \;, \la{Pi_larger}
\ee
and similarly for $1\to 2,3,4$.
The two Wightman functions are related by
\be
 \Pi^{<}_i(\omega,\vec{k}) = e^{-\omega/T}\, \Pi^{>}_i(\omega,\vec{k})
 \;, \la{rel1}
\ee
which is one way to see the origin of the Laplace transform
in \eq\nr{Laplace1}. 
The function $\Pi^{>}_i(\omega,\vec{k})$ in turn agrees with the spectral
function up to a trivial factor and exponentially small corrections, 
\be
 \Pi^{>}_i(\omega,\vec{k}) = 
 2 \bigl[ 1 + \nB{}(\omega) \bigr]\, \rho^{ }_i(\omega,\vec{k})
 \;, \la{rel2}
\ee
where $\nB{}$ is the Bose distribution. 

When the Wightman functions $\Pi^{>}_i$ corresponding to the operators in 
\eq\nr{L} are written in the basis of \eq\nr{def_H}, they have 
an overlap with many different ``elementary'' Wightman functions. 
The overlaps form a block-diagonal form, and can
be expressed through four different ``weight matrices'', 
denoted by $\mathcal{W}^{ }_i$: 
\ba
 \mathcal{W}^{ }_1 & \equiv & \quad 
 \begin{array}{cccc|c}
 \langle H^{ }_+ H^{ }_- & \langle H^{ }_0 H^{ }_0 & 
 \langle H^{ }_{\bar{0}} H^{ }_{\bar{0}} &
 \langle i H^{ }_{0} H^{ }_{\bar{0}} & \\[2mm] \hline
 & & & & \\[-4mm]
   \frac{4c^{ }_1 + c^{ }_2}{4} &
 \frac{4 c^{ }_1 - c^{ }_2}{8} & 
 \frac{4 c^{ }_1 - c^{ }_2}{8}
  &
  0 
  &
      H^\dagger_+ H^\dagger_- \rangle^{ }_{ } \\[3mm] 
 \frac{4 c^{ }_1 - c^{ }_2}{8} &
 \frac{c^{ }_2 + 4(c^{ }_1 + c^{ }_3 + c^{ }_4)}{16} &
 \frac{c^{ }_2 + 4(c^{ }_1 - c^{ }_3 - c^{ }_4)}{16} &
 \frac{c^{ }_3 - c^{ }_4}{2}
 &
      H^\dagger_0 H^\dagger_0 \rangle^{ }_{ } \\[3mm] 
 \frac{4 c^{ }_1 - c^{ }_2}{8} &
 \frac{c^{ }_2 + 4(c^{ }_1 - c^{ }_3 - c^{ }_4)}{16} &
 \frac{c^{ }_2 + 4(c^{ }_1 + c^{ }_3 + c^{ }_4)}{16} &
 \frac{c^{ }_4 - c^{ }_3}{2}
 &
      H^\dagger_{\bar{0}} H^\dagger_{\bar{0}} \rangle^{ }_{ } \\[3mm] 
 0 
   &
 \frac{c^{ }_3 - c^{ }_4}{2}
  &
 \frac{c^{ }_4 - c^{ }_3}{2}
 &
  c^{ }_3 + c^{ }_4
  &
      -i H^\dagger_0 H^\dagger_{\bar{0}} \rangle^{ }_{ }  
 \end{array}
 \quad \;, \quad \la{W1} \\[3mm]
 \mathcal{W}^{ }_2 & \equiv & \quad 
 \begin{array}{cc|c}
 \langle H^{ }_+ H^{ }_0 & 
 \langle i H^{ }_{+} H^{ }_{\bar{0}} & \\[2mm] \hline
 & & \\[-4mm] 
  \frac{c^{ }_2 + 4 c^{ }_4}{4}  &
  \frac{c^{ }_2 - 4 c^{ }_4}{4}  & 
      H^\dagger_+ H^\dagger_0 \rangle^{ }_{ } \\[3mm] 
 \frac{c^{ }_2 - 4 c^{ }_4}{4} &
 \frac{c^{ }_2 + 4 c^{ }_4}{4} &
      - i H^\dagger_+ H^\dagger_{\bar{0}} \rangle^{ }_{ } 
 \end{array}
 \quad \;, \quad \la{W2} \\[3mm]
 \mathcal{W}^{ }_3 & \equiv & \quad 
 \begin{array}{cc|c}
 \langle H^{ }_- H^{ }_0 & 
 \langle -i H^{ }_{-} H^{ }_{\bar{0}} & \\[2mm] \hline
 & & \\[-4mm] 
  \frac{c^{ }_2 + 4 c^{ }_3 }{4}  &
  \frac{c^{ }_2 - 4 c^{ }_3}{4}  & 
      H^\dagger_- H^\dagger_0 \rangle^{ }_{ } \\[3mm] 
 \frac{c^{ }_2 - 4 c^{ }_3}{4}  &
 \frac{c^{ }_2 + 4 c^{ }_3}{4}  &
      i H^\dagger_- H^\dagger_{\bar{0}} \rangle^{ }_{ } 
 \end{array}
 \quad \;, \quad \la{W3} \\[3mm]
 \mathcal{W}^{ }_4 & \equiv & \quad 
 \begin{array}{cc|c}
 \langle H^{ }_+ H^{ }_+ & 
 \langle H^{ }_{-} H^{ }_{-} & \\[2mm] \hline
 & & \\[-4mm] 
  c^{ }_4 &
  0 & 
      H^\dagger_+ H^\dagger_+ \rangle^{ }_{ } \\[3mm] 
  0 &
  c^{ }_3 &
      H^\dagger_- H^\dagger_- \rangle^{ }_{ } 
 \end{array}
 \quad \;. \quad \la{W4} 
\ea
Given that $c^{ }_3 = c^{ }_4$ (cf.\ \eq\nr{c3}), 
\eq\nr{W1} has itself a block-diagonal form. 

The right-hand sides of \eq\nr{Seq1}, which may be called the 
source terms, also turn into matrices in the basis of \eq\nr{def_H}.
These matrices are diagonal, but have in some cases non-trivial
coefficients, corresponding to the multiplicities of contractions:
\ba
 S^{ }_1(\vec{r},\vec{r}') & \equiv & 
 \begin{array}{cccc|c}
 \langle H^{ }_+ H^{ }_- & \langle H^{ }_0 H^{ }_0 & 
 \langle H^{ }_{\bar{0}} H^{ }_{\bar{0}} &
 \langle i H^{ }_{0} H^{ }_{\bar{0}} & \\[1mm] \hline 
 & & & & \\[-4mm]
  \delta^{(3)}(\vec{r-r'}) & 0 & 0 & 0 &
      H^\dagger_+ H^\dagger_- \rangle^{ }_{ } \\[1mm] 
  0 & 2\,\delta^{(3)}(\vec{r-r'}) & 0 & 0 &
      H^\dagger_0 H^\dagger_0 \rangle^{ }_{ } \\[1mm] 
  0 & 0 & 2\,\delta^{(3)}(\vec{r-r'}) & 0 &  
      H^\dagger_{\bar{0}} H^\dagger_{\bar{0}} \rangle^{ }_{ } \\[1mm] 
  0 & 0 & 0 &  \delta^{(3)}(\vec{r-r'}) &  
     -i H^\dagger_0 H^\dagger_{\bar{0}} \rangle^{ }_{ }  
 \end{array}
 \quad \;, \la{S1} \hspace*{9mm} \\[3mm]
  S^{ }_2(\vec{r},\vec{r}') & \equiv & 
 \begin{array}{cc|c}
 \langle H^{ }_+ H^{ }_0 & 
 \langle i H^{ }_{+} H^{ }_{\bar{0}} & \\[1mm] \hline 
 & & \\[-4mm]
 \delta^{(3)}(\vec{r-r'}) & 0  & 
      H^\dagger_+ H^\dagger_0 \rangle^{ }_{ } \\[1mm] 
 0  & \delta^{(3)}(\vec{r-r'}) &  
      - i H^\dagger_+ H^\dagger_{\bar{0}} \rangle^{ }_{ } 
 \end{array}
 \quad \;, \la{S2} \\[3mm]
  S^{ }_3(\vec{r},\vec{r}') & \equiv & 
 \begin{array}{cc|c}
 \langle H^{ }_- H^{ }_0 & 
 \langle -i H^{ }_{-} H^{ }_{\bar{0}} & \\[1mm] \hline 
 & & \\[-4mm]
 \delta^{(3)}(\vec{r-r'}) & 0  & 
      H^\dagger_- H^\dagger_0 \rangle^{ }_{ } \\[1mm] 
 0  & \delta^{(3)}(\vec{r-r'}) &  
      i H^\dagger_- H^\dagger_{\bar{0}} \rangle^{ }_{ } 
 \end{array}
 \quad \;, \la{S3} \\[3mm]
  S^{ }_4(\vec{r},\vec{r}') & \equiv & 
 \begin{array}{cc|c}
 \langle H^{ }_+ H^{ }_+ & 
 \langle H^{ }_{-} H^{ }_{-} &   \\[1mm] \hline 
 2\,\delta^{(3)}(\vec{r-r'}) & 0  & 
      H^\dagger_+ H^\dagger_+ \rangle^{ }_{ } \\[3mm] 
 0  & 2\,\delta^{(3)}(\vec{r-r'}) &  
      H^\dagger_- H^\dagger_- \rangle^{ }_{ } 
 \end{array}
 \quad \;. \la{S4}
\ea
As a crosscheck,  it may be noted that projecting the sources from 
\eqs\nr{S1}--\nr{S4} with 
the weights from \eqs\nr{W1}--\nr{W4} yields
\be
 \sum_{i=1}^{4} \tr \bigl[ \mathcal{W}^{ }_i\, S^{ }_i \bigr]
 \; = \;
 \Bigl[ 
   2 c^{ }_1 + \frac{3c^{ }_2}{2} + 6 (c^{ }_3 + c^{ }_4)
 \Bigr]\, \delta^{(3)}(\vec{r-r'}) 
 \;, 
\ee
which indeed agrees with weighted sum over the source terms
of \eq\nr{Seq1} with the normalization factors from \eq\nr{Ns}. 

The potentials can be derived as explained in ref.~\cite{threshold}, 
from the thermal expectation value of the time-evolution operator 
bracketed between states like in \eqs\nr{W1}--\nr{W4}. At this
point the sources are momentarily separated from each other; 
it is advantageous to symmetrize the 
state generated in this point-splitting, e.g.\ 
\be
 H^\dagger_+ H^\dagger_- \rightarrow
 H^\dagger_{\{+}(\vec{r}) H^\dagger_{-\}}(\vec{0})
 \equiv 
 \frac{1}{2} \bigl[ 
 H^\dagger_{+}(\vec{r}) H^\dagger_{-}(\vec{0})
 + 
 H^\dagger_{-}(\vec{r}) H^\dagger_{+}(\vec{0})
 \bigr]
 \;. 
\ee
Then a straightforward computation produces matrix potentials, 
listed in table~\ref{table:potentials}.

%
\begin{table}[ph]

\small{
\begin{center}
\begin{tabular}{cccc|c}
 \multicolumn{5}{c}{
 $\mathcal{U}^{ }_1(r)$
 }
 \\[3mm] \hline\hline & & & & \\[-1mm]
 $\langle H^{ }_+ H^{ }_-$ &
 $\langle H^{ }_0 H^{ }_0$ & 
 $\langle H^{ }_{\bar{0}} H^{ }_{\bar{0}}$ &
 $\langle i H^{ }_{0} H^{ }_{\bar{0}}$ &
 \\[2mm] \hline & & & & \\[-1mm]
 $\mathcal{V}^{ }_\rmiii{\bar{Z}\bar{Z}}(0) 
  + 2\mathcal{V}^{ }_\rmiii{{W}{W}}(0)
  -\mathcal{V}^{ }_\rmiii{\bar{Z}\bar{Z}}(r)$
    \hspace*{-5mm}
  &
 $-\mathcal{V}^{ }_\rmiii{{W}{W}}(r)$  &
 $-\mathcal{V}^{ }_\rmiii{{W}{W}}(r)$  &
  0   &
 $H^\dagger_+ H^\dagger_- \rangle^{ }_{ }$  \\[3mm] 
 $ -2\,\mathcal{V}^{ }_\rmiii{{W}{W}}(r)$  &
    \hspace*{-3mm}
 $  \mathcal{V}^{ }_\rmiii{{Z}{Z}}(0) 
  + 2\mathcal{V}^{ }_\rmiii{{W}{W}}(0)$
    \hspace*{-3mm}
    & 
 $ - \mathcal{V}^{ }_\rmiii{{Z}{Z}}(r)$  &
   0  &
 $   H^\dagger_0 H^\dagger_0 \rangle^{ }_{ }$  \\[3mm] 
 $ -2\,\mathcal{V}^{ }_\rmiii{{W}{W}}(r)$  &
 $ - \mathcal{V}^{ }_\rmiii{{Z}{Z}}(r)$  &
    \hspace*{-3mm}
 $   \mathcal{V}^{ }_\rmiii{{Z}{Z}}(0) 
  + 2\mathcal{V}^{ }_\rmiii{{W}{W}}(0)$
    \hspace*{-3mm}
   & 
  0 & 
 $   H^\dagger_{\bar{0}} H^\dagger_{\bar{0}} \rangle^{ }_{ }$  \\[3mm] 
 0 
   &
 0 
  &
 0 
  & \hspace*{-5mm}
 $  \mathcal{V}^{ }_\rmiii{{Z}{Z}}(0) 
  + 2\mathcal{V}^{ }_\rmiii{{W}{W}}(0)
  + \mathcal{V}^{ }_\rmiii{{Z}{Z}}(r)$
  &
 $   -i H^\dagger_0 H^\dagger_{\bar{0}} \rangle^{ }_{ }$  
\end{tabular}

\vspace*{6mm}

\begin{tabular}{cc|c}
 \multicolumn{3}{c}{
 $\mathcal{U}^{ }_2(r)$
 }
 \\[3mm] \hline\hline & & \\[-1mm]
 $\langle H^{ }_+ H^{ }_0$ & 
 $\langle i H^{ }_{+} H^{ }_{\bar{0}}$ &
 \\[2mm] \hline & & \\[-1mm] 
    $\frac{1}{2}\bigl[ \mathcal{V}^{ }_\rmiii{{Z}{Z}}(0)
  + \mathcal{V}^{ }_\rmiii{\bar{Z}\bar{Z}}(0)\bigr]
  + 2\mathcal{V}^{ }_\rmiii{{W}{W}}(0)
  + \mathcal{V}^{ }_\rmiii{{W}{W}}(r)$  &
  $- \mathcal{V}^{ }_\rmiii{{W}{W}}(r)
  - \mathcal{V}^{ }_\rmiii{{Z}\bar{Z}}(r)$  &
      $H^\dagger_+ H^\dagger_0 \rangle^{ }_{ }$ \\[3mm] 
  $- \mathcal{V}^{ }_\rmiii{{W}{W}}(r)
  - \mathcal{V}^{ }_\rmiii{{Z}\bar{Z}}(r)$  &
  $\frac{1}{2} \bigl[ \mathcal{V}^{ }_\rmiii{{Z}{Z}}(0)
  + \mathcal{V}^{ }_\rmiii{\bar{Z}\bar{Z}}(0)\bigr]
  + 2\mathcal{V}^{ }_\rmiii{{W}{W}}(0)
  + \mathcal{V}^{ }_\rmiii{{W}{W}}(r)$  &
  $  - i H^\dagger_+ H^\dagger_{\bar{0}} \rangle^{ }_{ }$ 
\end{tabular}

\vspace*{6mm}

\begin{tabular}{cc|c}
 \multicolumn{3}{c}{
 $\mathcal{U}^{ }_3(r)$
 }
 \\[3mm] \hline\hline & & \\[-1mm]
 $\langle H^{ }_- H^{ }_0$ & 
 $\langle -i H^{ }_{-} H^{ }_{\bar{0}}$ & \\[2mm] \hline
 & & \\[-1mm] 
 $ \frac{1}{2}\bigl[ \mathcal{V}^{ }_\rmiii{{Z}{Z}}(0)
  + \mathcal{V}^{ }_\rmiii{\bar{Z}\bar{Z}}(0)\bigr]  
  + 2\mathcal{V}^{ }_\rmiii{{W}{W}}(0)
  + \mathcal{V}^{ }_\rmiii{{W}{W}}(r)$  &
 $ - \mathcal{V}^{ }_\rmiii{{W}{W}}(r)
  - \mathcal{V}^{ }_\rmiii{{Z}\bar{Z}}(r)$  &
 $     H^\dagger_- H^\dagger_0 \rangle^{ }_{ }$ \\[3mm] 
 $ - \mathcal{V}^{ }_\rmiii{{W}{W}}(r)
  - \mathcal{V}^{ }_\rmiii{{Z}\bar{Z}}(r)$  &
 $  \frac{1}{2} \bigl[ \mathcal{V}^{ }_\rmiii{{Z}{Z}}(0)
  + \mathcal{V}^{ }_\rmiii{\bar{Z}\bar{Z}}(0)\bigr]  
  + 2\mathcal{V}^{ }_\rmiii{{W}{W}}(0)
  + \mathcal{V}^{ }_\rmiii{{W}{W}}(r)$  &
 $     i H^\dagger_- H^\dagger_{\bar{0}} \rangle^{ }_{ }$ 
\end{tabular}

\vspace*{6mm}

\begin{tabular}{cc|c}
 \multicolumn{3}{c}{
 $\mathcal{U}^{ }_4(r)$
 } 
 \\[3mm] \hline\hline & & \\[-1mm]
 $\langle H^{ }_+ H^{ }_+$ & 
 $\langle H^{ }_{-} H^{ }_{-}$ & 
 \\[2mm] \hline & & \\[-1mm] 
 $
  \mathcal{V}^{ }_\rmiii{\bar{Z}\bar{Z}}(0)
  + 2\mathcal{V}^{ }_\rmiii{{W}{W}}(0)
  + \mathcal{V}^{ }_\rmiii{\bar{Z}\bar{Z}}(r)$ &
  0 & 
 $     H^\dagger_+ H^\dagger_+ \rangle^{ }_{ }$ \\[3mm] 
  0 &
 $ \mathcal{V}^{ }_\rmiii{\bar{Z}\bar{Z}}(0)
  + 2\mathcal{V}^{ }_\rmiii{{W}{W}}(0)
 + \mathcal{V}^{ }_\rmiii{\bar{Z}\bar{Z}}(r)$ &
 $     H^\dagger_- H^\dagger_- \rangle^{ }_{ }$ 
\end{tabular}
\end{center}
}

\vspace*{3mm}

\caption[a]{\small
  The ``potentials'' $\mathcal{U}^{ }_i$
  appearing in \eq\nr{Seq}.  In general the potentials contain
  both a real part, as well as
  an imaginary part representing thermal scatterings 
  (cf.\ \eq\nr{W0_prop} and \se\ref{ss:applicable}).
 }

\label{table:potentials}
\end{table}
%

Apart from \eq\nr{V_WW}, 
the potentials in table~\ref{table:potentials} contain the object
\be
 \mathcal{V}^{ }_\rmii{${Z}\bar{Z}$}(r) \; \equiv \; 
 \frac{\tilde{g}^2}{4}
 \int_{\vec{k}} e^{i \vec{k}\cdot\vec{r}}
 \, i \langle Z^{ }_0 \bar{Z}^{ }_0 \rangle^{ }_\rmii{T}(0,\vec{k})
 \;, \la{VZZ}
\ee
and similarly for 
$ 
 \mathcal{V}^{ }_\rmii{${Z}{Z}$}
$
and 
$
 \mathcal{V}^{ }_\rmii{$\bar{Z}\bar{Z}$}
$,
where we have defined 
\be
 \tilde{g} Z^{ }_0 \; \equiv \; g^{ }_1 B^{ }_0 + g^{ }_2 A^3_0
 \;,  \quad
 \tilde{g} \bar{Z}^{ }_0 \; \equiv \; g^{ }_1 B^{ }_0 - g^{ }_2 A^3_0
 \;, \quad
 \tilde{g} \; \equiv \; \sqrt{g_1^2 + g_2^2}
 \;. \la{redef}
\ee
We stress that at finite temperature $Z^{ }_0$ does not represent
a propagating mode, and $\bar{Z}^{ }_0$ does not represent one even
at zero temperature. The fields $Z^{ }_0$ and $\bar{Z}^{ }_0$ simply
stand for specific linear combinations originating from vertices; 
the diagonal modes are obtained from $B^{ }_0$ and $A^3_0$ through
an orthogonal transformation parametrized by 
a temperature-dependent mixing angle $\tilde\theta$, given in 
\eq\nr{mixing}. 

The potentials of table~\ref{table:potentials}
contain a real part, including the diagonal
$r$-independent Salpeter correction, as well as an imaginary part, 
representing scatterings and decays of the type described
by \eq\nr{reactions}. As mentioned in \se\ref{ss:applicable}, 
the inclusion of the scatterings has been demonstrated 
to be theoretically consistent in the case of the most attractive
channel, in which case the scattering rate is 
a slow one. This slow rate appears in the upper diagonal
block of the potential $\mathcal{U}^{ }_1$ in 
table~\ref{table:potentials}. Its role is to damp 
(or ``decohere'') 
oscillations between the three states appearing in this block.
In the other channels, the widths represent a part of NLO corrections. 

With these ingredients at hand, 
the thermally averaged scattering rates
are obtained from matrix Schr\"odinger equations of the form
\ba
 \biggl[ 
   -\frac{\nabla_r^2}{M} + \mathop{\mbox{diag}}(\Delta M)
   + \mathcal{U}^{ }_i(r) - E'
 \biggr] F^{ }_i(E';\vec{r},\vec{r'}) & = & 
 S^{ }_i(\vec{r},\vec{r'}) \quad \mbox{(no sum over $i$)}\;, \la{Seq} \\ 
 \lim_{\vec{r,r'}\to \vec{0}} \im  F^{ }_i(E';\vec{r},\vec{r'})
 & = & \varrho^{ }_i(E') \la{get_rho}
 \;,
\ea
where the matrix $F^{ }_i$ has the same dimension as the source $S^{ }_i$.
The combination needed for \eq\nr{sigma} becomes, in analogy with 
\eq\nr{Laplace1}, 
\ba
 \sum_{i=1}^4 c^{ }_i \gamma^{ }_i & \approx &
 \Bigl( \frac{MT}{\pi} \Bigr)^{3/2} e^{-2M/T}
 \int_{- \Lambda}^{\infty} 
 \! \frac{{\rm d}E'}{\pi} \, e^{-E'/T}
 \; \sum_{i=1}^4 \tr \bigl[ \mathcal{W}^{ }_i\, \varrho^{ }_i(E') \bigr]
 \;. \la{Laplace2}
\ea 


It is interesting to ask how the degenerate limit of \se\ref{ss:degenerate}
is recovered from the equations of the current section. A simple way to do
this is to recall that if a Green's function is expressed as a function
of time $t$ rather than energy $E'$, then the source terms in 
\eqs\nr{S1}--\nr{S4} represent initial conditions at time $t=0$~\cite{peskin}. 
To first order in interactions, we can simply act on the initial conditions
with the potentials of table~\ref{table:potentials}, and subsequently project
the results with the weights from \eqs\nr{W1}--\nr{W4}, i.e.\ 
compute 
$
 \sum_{i=1}^4 \tr \bigl[ \mathcal{W}^{ }_i\, \mathcal{U}^{ }_i S^{ }_i \bigr]
$. 
It can be verified that the terms proportional to  
$
  2 c^{ }_1\, \delta^{(3)}(\vec{r-r'})
$, 
$
  \frac{3c^{ }_2}{2}\, \delta^{(3)}(\vec{r-r'})
$, 
and 
$
   6\, ( c^{ }_3 + c^{ }_4  )\, \delta^{(3)}(\vec{r-r'})
$
reproduce the potentials from 
\eqs\nr{pot1}, \nr{pot2} and \nr{pot3}, respectively. 

%
\subsection{Limit of low temperatures}
\la{ss:lowT}

The scale hierarchy shown in \eq\nr{scales} breaks down 
as the temperature decreases: 
first the Debye scale $gT$ becomes smaller than the energy
scale $\alpha^2 M$ at which the Schr\"odinger description
applies, and soon afterwards $\pi T$ also becomes smaller
than $\alpha^2 M$. Moreover, assuming that mass splittings 
in the dark sector are $\Delta M^{ }_{ij} \gsim \alpha^2 M$, $\pi T$ also 
becomes smaller than $\Delta M^{ }_{ij}$. These crossings 
have an important impact on the determination of 
$\sum_i c^{ }_i \gamma^{ }_i$ and $\langle \sigma^{ }_\rmi{eff}\, v \rangle$
at low temperatures, particularly as far as the  
below-threshold part ($E' < 0$) is concerned, 
given that the Laplace transforms
in \eqs\nr{Laplace1} and \nr{Laplace2} exponentially enhance 
the contributions from the smallest energies.

%
\begin{figure}[t]
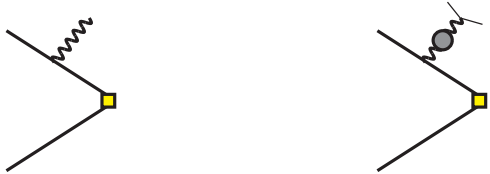


\hspace*{6mm}%
\begin{minipage}[c]{14.2cm}
\begin{eqnarray*}
&& 
 \hspace*{-3.5cm}
 \NRa 
 \hspace*{0.5cm}
 \NRb
\end{eqnarray*}
\end{minipage}

\vspace*{4mm}

\caption[a]{\small 
 Left: absorption or emission
 of an on-shell gauge boson
 by an annihilating dark matter particle. 
 Right: $t$-channel scattering of a dark matter particle
 off thermal Standard Model particles, denoted by thin lines. The filled blob 
 indicates that, due to infrared sensitivity,
 the soft gauge boson has to dressed by thermal 
 corrections such as Debye screening. The process on the right
 dominates in the range of \eq\nr{scales}, 
 whereas the process on the left dominates
 at low temperatures (cf.\ \se\ref{ss:lowT}). 
 }
\la{fig:processes}
\end{figure}
%

It may be noted, first of all, that once
the Debye scale drops below $\alpha^2 M$, 
the dominant process responsible for the thermal 
interaction rate is the absorption of a thermal gauge 
boson (cf.\ \fig\ref{fig:processes}(left)) rather 
than scattering off Standard Model particles as is 
the case at higher temperatures
(cf.\ \fig\ref{fig:processes}(right))
(cf.\ ref.~\cite{jacopo2} and references therein). 
However, this does not change the magnitude of the 
thermal interaction rate qualitatively. Given that
the numerical effect from the low-temperature regime
is modest, we have not worked out these effects
quantitatively; this would
pose an interesting topic for future research. 

More importantly, the spectral function changes dramatically
once $\pi T \lsim |E'| \sim \alpha^2 M$. In vacuum, 
the spectral function vanishes for $E' < 0$ in repulsive
channels, and for $E'$ below the ground state energy in attractive
channels. At $T> 0$, this is no longer the case: any ``measurement''
can detect non-vanishing below-threshold spectral weight, with the
energy difference to the vacuum threshold 
supplied by a thermal fluctuation suppressed 
by a Boltzmann factor. This has 
been shown explicitly in a QCD context, both by considering
the dissociation rate of bound states with pNRQCD 
(cf.\ \eq(89) of ref.~\cite{jacopo}), and through 
a strict NLO computation of the process
in \fig\ref{fig:processes}(left) together with the associated
virtual corrections 
(cf.\ \eq(4.7) of ref.~\cite{nlo} after setting $\omega\to 2 M + \Delta E'$).  
We have not carried out a quantitative analysis of these effects
for the present system, which
would again pose an interesting topic for future research, however
we multiply thermal interaction rates by the Boltzmann factor
$\theta(-E')e^{-|E'|/T}$ in order to account for 
the exponential suppression below threshold. This is a higher-order 
effect in the domain of our main interest, \eq\nr{scales}, but  
imposes the correct overall magnitude to the below-threshold
spectral function when $\pi T \lsim \alpha^2 M$. 

The third effect concerns mass splittings, which are
always present at least at the level 
$\Delta M^{ }_{ij} \sim 10^{-3}M$~\cite{minimal}. 
To account for them 
properly requires the numerical solution of the matrix equations
derived in \se\ref{ss:nondeg}. However, on the qualitative level
we can profit from a corresponding solution that was worked out in 
sec.~7 of ref.~\cite{threshold}. The main finding was that
as long as $\Delta M^{ }_{ij} \sim \alpha^2 M$, the shape of the 
spectral function does not depend noticeably on $\Delta M^{ }_{ij}$, 
however the spectral function 
splits into several parts, separated by the mass shifts.\footnote{%
 The shape stays intact because the heavier particles still contribute
 as virtual states and thereby generate an interaction between
 the lightest ones. 
 } 
We can work out these shifts by solving \eqs\nr{Seq} and \nr{get_rho}
at tree level but with $\Delta M^{ }_{ij}\neq 0$.\footnote{%
 Thermal mass corrections can be omitted in this regime, 
 given that $|\Delta M^{ }_\rmii{$T$}| \sim \alpha^{3/2} T 
 \lsim \alpha^{7/2} M \ll \alpha^2 M$.
 } 
Denoting by 
$\rho^{(0)}_{ } \equiv 
M^{\fr32}\theta(E') \sqrt{E'}/(4\pi)$ 
the tree-level spectral function obtained
with $\Delta M^{ }_{ij} = 0$, and using
$\Delta M^{ }_+ = \Delta M^{ }_-$ 
(here 
$\Delta M^{ }_{i_1..i_n} \equiv M^{ }_{i_1} + ... +  M^{ }_{i_n} - n M$), 
we find
\ba
 \sum_{i=1}^4 \tr \bigl[ \mathcal{W}^{ }_i\, \varrho^{(0)}_i(E') \bigr]
 & = &  
 2 c^{ }_1 \,  
 \Biggl[ 
  \frac{\rho^{(0)}_{ }(E')}{4}
 + 
   \frac{\rho^{(0)}_{ }(E' - \Delta M^{ }_{\bar{0}\bar{0}})
   + 2 \rho^{(0)}_{ }( E' - \Delta M^{ }_{+-} )  }{4}
 \Biggr]
 \nn 
 & + & 
 \frac{3c^{ }_2}{2} \,  
 \Biggl[ 
  \frac{\rho^{(0)}_{ }(E')}{12}
 + 
  \frac{\rho^{(0)}_{ }(E' - \Delta M^{ }_+)}{3}
 \nn 
 & + & \hspace*{4mm}
   \frac{
       \rho^{(0)}_{ }(E' - \Delta M^{ }_{\bar{0}\bar{0}})
   + 4 \rho^{(0)}_{ }(E' - \Delta M^{ }_{\bar{0}+})
   + 2 \rho^{(0)}_{ }( E' - \Delta M^{ }_{+-} ) 
  }{12}
 \Biggr]
 \nn 
 & + & 
 6 ( c^{ }_3 + c^{ }_4 ) \,  
 \Biggl[ 
  \frac{\rho^{(0)}_{ }(E')}{12}
 + 
  \frac{\rho^{(0)}_{ }(E' - \Delta M^{ }_{\bar{0}})
       + \rho^{(0)}_{ }(E' - \Delta M^{ }_+) }{6}
 \nn 
 & + & \hspace*{4mm}
   \frac{
       \rho^{(0)}_{ }(E' - \Delta M^{ }_{\bar{0}\bar{0}})
   + 2 \rho^{(0)}_{ }(E' - \Delta M^{ }_{\bar{0}+})
   + 4 \rho^{(0)}_{ }( E' - \Delta M^{ }_{+-} ) 
  }{12}
 \Biggr]
 \;. \nn[2mm] 
 \la{DeltaMij}
\ea 
Inserting this into \eq\nr{Laplace2}, the contributions
of the shifted thresholds get suppressed by $e^{-\Delta M^{ }_{ij}/T}$
just like in \eq\nr{sigma_eff}; 
a practical implementation is shown in 
\eqs\nr{barS_thres1} and \nr{barS_thres2}. 

%
\section{Numerical solution and overclosure bound}
\la{se:numerics}

Once the combination $\sum_i c^{ }_i\gamma^{ }_i$ has been computed as 
a function of the temperature, either from \eq\nr{Laplace1} or 
from \eq\nr{Laplace2}, the effective 
cross section $\langle \sigma^{ }_\rmi{eff} v \rangle$ is obtained 
from \eq\nr{sigma}. Writing out the time derivative in \eq\nr{boltzmann}, 
the evolution equation reads
\be
 (\partial^{ }_t + 3 H ) n = - 
 \langle \sigma^{ }_\rmi{eff} v^{ } \rangle \, (n^2 - n_\rmi{eq}^2)
 \;, 
\ee
where $H$ is the Hubble rate. Combining this with the entropy 
conservation law
$
 (\partial^{ }_t + 3 H ) s = 0 
$
as well as with the relation of time and temperature, 
$
 \dot{T} =  - {3 H s}/{c} 
$,
where $c$ is the heat capacity; defining  
a ``yield parameter'' through $Y \equiv n/s$;
and denoting $z \equiv M/T$, we get 
\be
 Y'(z) 
   = - \, 
   \langle \sigma^{ }_\rmi{eff}\, v \rangle M m^{ }_\rmi{Pl} \times 
   \frac{c(T)}{\sqrt{24\pi e(T)}} \times
   \left. \frac{Y^2(z) - Y^{2}_\rmi{eq}(z) }{z^2} \right|^{ }_{T = M/z}
 \;. \la{dY}
\ee
Here $m^{ }_\rmi{Pl}$ is the Planck mass and $e$ is the energy density. 
We insert $e$, $c$, and $s$ from ref.~\cite{crossover}.

Our goal is to determine 
a conservative overclosure bound for $M$. Thus, for a 
given $M$, we need a lower bound for $Y$. A lower bound for $Y$
requires an upper bound for 
$
 \langle \sigma^{ }_\rmi{eff}\, v \rangle
$,
so that annihilations take place with maximal efficiency. 
As discussed in \se\ref{ss:lowT}, 
if $\Delta M^{ }_{ij} \sim \alpha^2 M \sim 10^{-3} M$, 
then in the 
non-degenerate situation the solution of the Schr\"odinger equation
does not differ qualitatively from the degenerate limit. 
In fact  
$
 \langle \sigma^{ }_\rmi{eff}\, v \rangle
$
decreases with $\Delta M^{ }_{ij}$, because of the 
Boltzmann suppression factors $\sim e^{-\Delta M^{ }_{ij}/T}$ induced
by the movement of the heavier particle thresholds to higher energies. 
Therefore, the degenerate
limit sets an upper bound for 
$
 \langle \sigma^{ }_\rmi{eff}\, v \rangle
$. 
We only depart from this approximation at very low temperatures
$\pi T \lsim \alpha^2 M$ where effects from $\Delta M^{ }_{ij}$
start to be of order unity (cf.\ \eqs\nr{barS_thres1} and \nr{barS_thres2}).

For numerical evaluations, the gauge couplings $g_1^2$ and $g_2^2$, 
the top Yukawa coupling $h_t^2$, 
and the scalar couplings appearing in \eq\nr{V} are needed. 
The gauge couplings affecting the ``soft'' thermal physics of the static 
potential are evaluated at a scale $\bmu \simeq \pi T$.
In contrast the couplings in \eqs\nr{c1}--\nr{c3} are needed 
at a scale $\bmu \simeq 2 M$. We fix 
$g_1^2(\mZ^{ }) = 0.128$, $g_2^2(\mZ^{ }) = 0.425$, 
$h_t^2(\mZ^{ }) = 0.967$, $\lambda^{ }_1(\mZ^{ }) = 0.145$, 
and for $\bmu < \mZ$ keep these unchanged. 
For $\mZ^{ } < \bmu < M$, 
the couplings are evolved like in the Standard Model, e.g.\ 
$
 g_1^2(\bmu) \approx 48\pi^2/ [41 \ln(\Lambda^{ }_1/\bmu)]
$
and
$
 g_2^2(\bmu) \approx 48\pi^2/ [19 \ln(\bmu / \Lambda^{ }_2)] 
$.   
For $\bmu > M$ we switch to the IDM evolution~\cite{idm_rg}, 
$
 g_1^2(\bmu) \approx 48\pi^2/ [42 \ln(\Lambda'_1/\bmu)]
$
and
$
 g_2^2(\bmu) \approx 48\pi^2/ [18 \ln(\bmu / \Lambda'_2)] 
$.   

\begin{figure}[t]
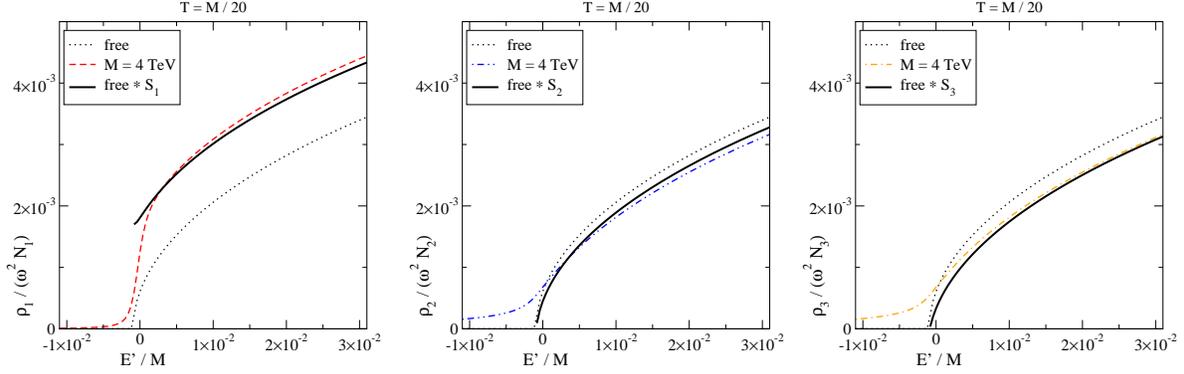


\hspace*{-0.1cm}
\centerline{%
 \epsfxsize=5.0cm\epsfbox{rho_M4000_V1.eps}
 \hspace{0.1cm}
 \epsfxsize=5.0cm\epsfbox{rho_M4000_V2.eps}%
 \hspace{0.1cm}
 \epsfxsize=5.0cm\epsfbox{rho_M4000_V3.eps}%
}

\caption[a]{\small
 The free (dotted lines; shifted
 by the Salpeter correction) and resummed
 (coloured lines; cf.\ \eq\nr{get_rho1})  spectral functions for 
 $M=4$~TeV, $T = M/20$, compared
 with results predicted by the massless Sommerfeld factors 
 (solid lines; cf.\ \eq\nr{S}). The potentials are from 
 \eqs\nr{pot1}--\nr{pot3}; $\mathcal{V}^{ }_1$ is 
 attractive and $\mathcal{V}^{ }_{2,3}$ are repulsive at short
 distances ($\mathcal{V}^{ }_{2}$ is attractive 
 at large distances). The spectral function $\rho^{ }_1$ obtains a
 more dramatic shape at low temperatures, cf.\ \fig\ref{fig:Omega}(left). 
}

\la{fig:Z_M2}
\end{figure}

Examples of spectral functions from \eq\nr{get_rho1}, for the three 
potentials from \eqs\nr{pot1}--\nr{pot3}, are shown in \fig\ref{fig:Z_M2}
for $M = 4$~TeV, $T = M/20$.\footnote{%
 For the numerical solution we employ the same method as
 in ref.~\cite{threshold}, originally introduced in ref.~\cite{original}. 
 } 
The results are compared with 
massless Sommerfeld factors from \eq\nr{S}, shifted
by the Salpeter correction in \eq\nr{salpeter}. 
Reasonable agreement is found, in spite of the presence of 
Debye screening and complicated mixing patterns that appear 
in the thermal potentials. 

Consider now 
$ 
 \langle \sigma^{ }_\rmi{eff}\, v \rangle 
$ from \eq\nr{sigma}. 
It is convenient to express the result in a form similar 
to \eq\nr{sigma_free}, 
\be
 \langle \sigma^{ }_\rmi{eff}\, v \rangle^{ }_{ } \; = \; 
 \frac{c^{ }_1 \bar{S}^{ }_1}{2} 
 + \frac{3 c^{ }_2 \bar{S}^{ }_2 }{8} 
 + \frac{3(c^{ }_3 + c^{ }_4) \bar{S}^{ }_3}{2} 
 \;, \la{sigma_full}
\ee
where ``average Sommerfeld factors'' have been defined as 
\be
 \bar{S}^{ }_i \; \equiv \; 
 \frac{  e^{2 \Delta M^{ }_\rmii{$T$} / T} }{N^{ }_i} 
 \biggl( \frac{4 \pi}{MT} \biggr)^{\fr32}
 \int_{-\Lambda}^{\infty} \! \frac{{\rm d}E'}{\pi}
 e^{-{E'} / {T}} \rho^{ }_i(E') 
 \;. \la{barS}
\ee
The Salpeter correction is given by \eqs\nr{pot1}--\nr{pot3}, 
\nr{VWW0}, \nr{VAA0}, and \nr{VBB0}, 
\be
 2 \Delta M^{ }_\rmii{$T$} \; \equiv \; \re
  \bigl[ \, 
     2 \mathcal{V}^{ }_\rmii{${W}{W}$}(0)
   + \mathcal{V}^{ }_\rmii{${A}{A}$}(0)  
   + \mathcal{V}^{ }_\rmii{${B}{B}$}(0)
  \, \bigr]
 \;. \la{salpeter}
\ee
Its appearance in \eq\nr{barS} 
originates from the fact that $1/n_\rmi{eq}^2$ in 
\eq\nr{sigma} gets changed, 
\be
 n^{ }_\rmi{eq} \approx 4\,  
 \biggl( \frac{MT}{2 \pi} \biggr)^{\fr32} e^{-(M + \Delta M^{ }_\rmii{$T$})/T}
 \;. \la{neq2}
\ee
If the change of the threshold location were the only modification
of the spectral function $\rho^{ }_i$, $2\Delta M^{ }_\rmii{$T$}$ 
would exactly cancel out in \eq\nr{barS}.

As discussed in \se\ref{ss:lowT},  
the vacuum mass differences $\Delta M^{ }_{ij}$  
become important at very low temperatures
(in contrast $\Delta M^{ }_\rmii{$T$}$ loses its significance there).
Inserting \eq\nr{DeltaMij} into \eq\nr{Laplace2}, 
comparing with \eq\nr{sigma_full}, and setting for simplicity
$\Delta M^{ }_+ = \Delta M^{ }_{\bar{0}} \equiv \Delta M$, 
the effects from $\Delta M$ can phenomenologically be included
through the substitutions
\ba
 \bar{S}^{ }_1 & \to & 
 \bar{S}^{ }_{1,\rmi{eff}} \; \equiv \; 
 \bar{S}^{ }_1 \, \Biggl[ 
   \frac{1}{4} + \frac{3e^{-2 \Delta M/T}}{4}
 \Biggr]
 \;, \la{barS_thres1} \\ 
 \bar{S}^{ }_{2,3,4} & \to & 
 \bar{S}^{ }_{2,3,4,\rmi{eff}} \; \equiv \; 
 \bar{S}^{ }_{2,3,4} \, \Biggl[ 
   \frac{1}{12}
  +  \frac{e^{- \Delta M/T}}{3}
  +  \frac{7e^{-2 \Delta M/T}}{12}
 \Biggr]
 \;. \la{barS_thres2}
\ea
We adopt this recipe in the following, setting for illustration
$\Delta M = 10^{-3} M$, 
which is parametrically in the correct range $\sim \alpha^2 M$
and numerically in fair accordance with ref.~\cite{minimal} at
$\lambda^{ }_i = 0$, and also reflects
the gradual increase of $\Delta M \simeq \lambda^{ }_{4,5}v^2 / M$
with scalar self-couplings. The case $\Delta M = 0$ is considered
as an upper bound on the average Sommerfeld factors. 

The average Sommerfeld factors have been plotted in \fig\ref{fig:barS}.
For the numerical evaluation of \eq\nr{barS}, we have restricted 
the Laplace transform to the range $E'\in (E'_\rmi{min},E'_\rmi{max})$, 
where $E'_\rmi{min} \equiv 2 \Delta M^{ }_\rmii{$T$} - 15 \alpha^2 M$ and 
$E'_\rmi{max} \equiv 15 T$, where $\alpha\equiv (g_1^2 + 3g_2^2)/(16\pi)$.

\begin{figure}[t]

\hspace*{-0.1cm}
\centerline{%
 \epsfxsize=5.0cm\epsfbox{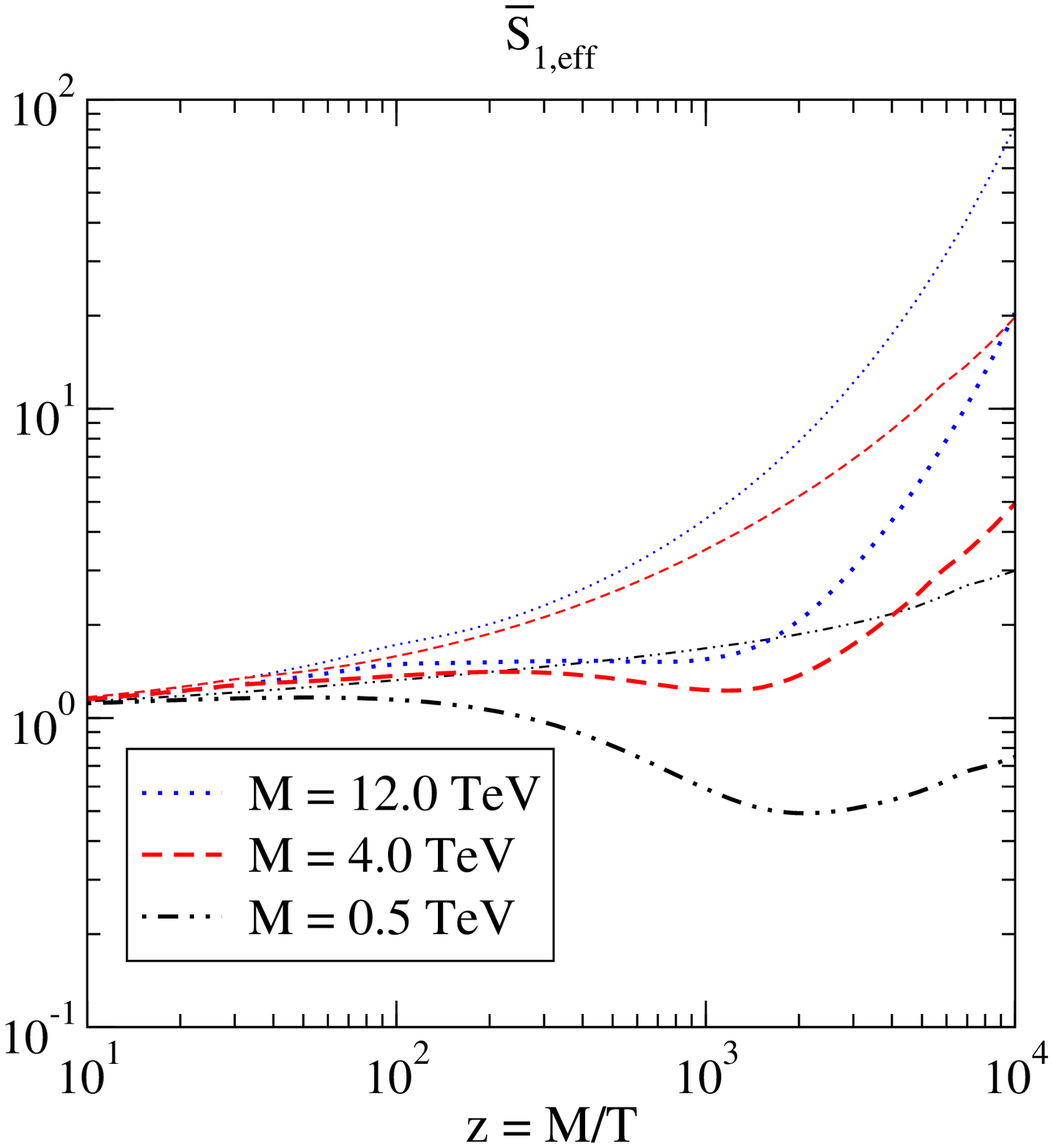}
 \hspace{0.1cm}
 \epsfxsize=5.0cm\epsfbox{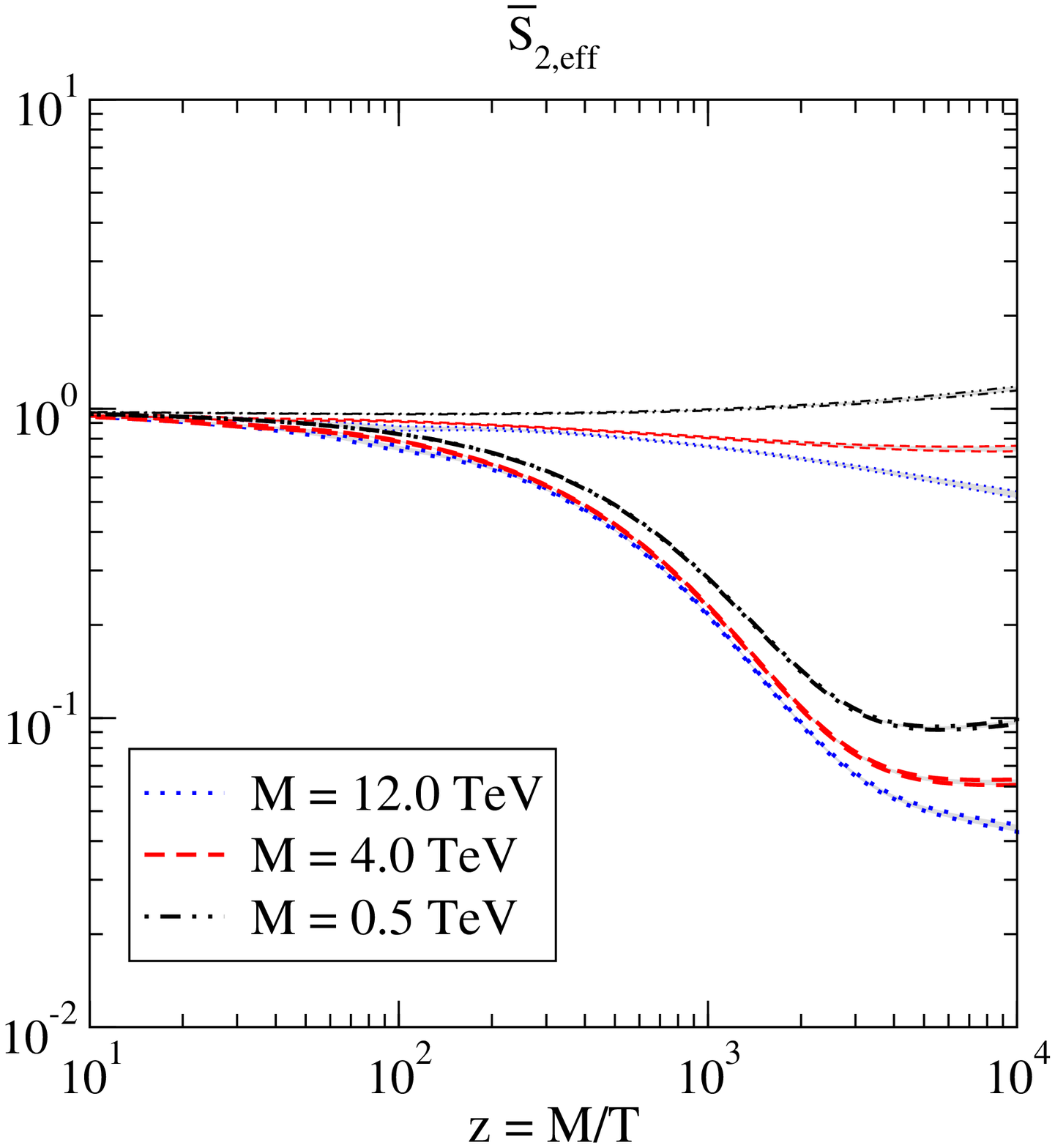}%
 \hspace{0.1cm}
 \epsfxsize=5.0cm\epsfbox{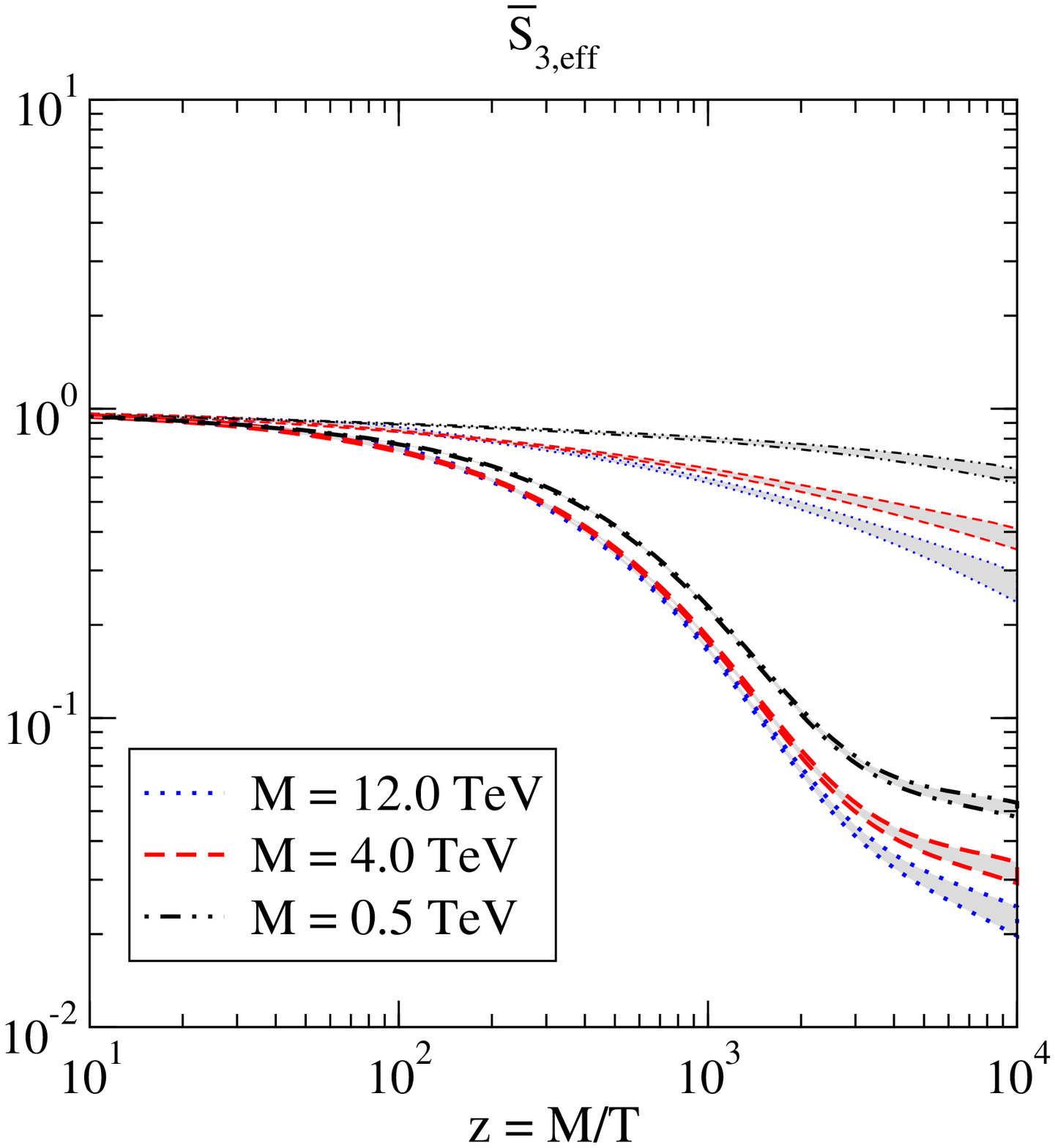}%
}

\caption[a]{\small
 Thin lines show 
 the average Sommerfeld factors from \eq\nr{barS},
 as a function of $z\equiv M/T$, for $\Delta M = 0$; thick lines
 include the modifications
 according to \eqs\nr{barS_thres1} and \nr{barS_thres2},
 with $\Delta M \equiv 10^{-3} M$.
 For $\bar{S}^{ }_{2,3}$ the error band indicates the uncertainty
 as discussed around the end of \se\ref{ss:applicable}.  
 For large $z$, $\bar{S}^{ }_1$ increases because of 
 the emergence of bound-state like structures just
 below threshold (cf.\ \fig\ref{fig:Omega}(left)).
}

\la{fig:barS}
\end{figure}

Given the average Sommerfeld factors, we can insert \eq\nr{sigma_full}
into \eq\nr{dY} and integrate the latter equation for $Y(z)$. 
Examples of solutions are shown in \fig\ref{fig:Y}. We have  
compared with the linearized version of this equation (cf.\ \eq\nr{response}), 
obtained by setting 
$
 Y^2 - Y_\rmi{eq}^2 \to 2 Y^{ }_\rmi{eq}(Y - Y^{ }_\rmi{eq})
$. It is observed how the initial
departure from equilibrium is well described by both forms, 
however afterwards the Lee-Weinberg from of \eq\nr{dY} leads to 
a substantial depletion of the dark matter abundance. 

\begin{figure}[t]

\hspace*{-0.1cm}
\centerline{%
 \epsfxsize=5.0cm\epsfbox{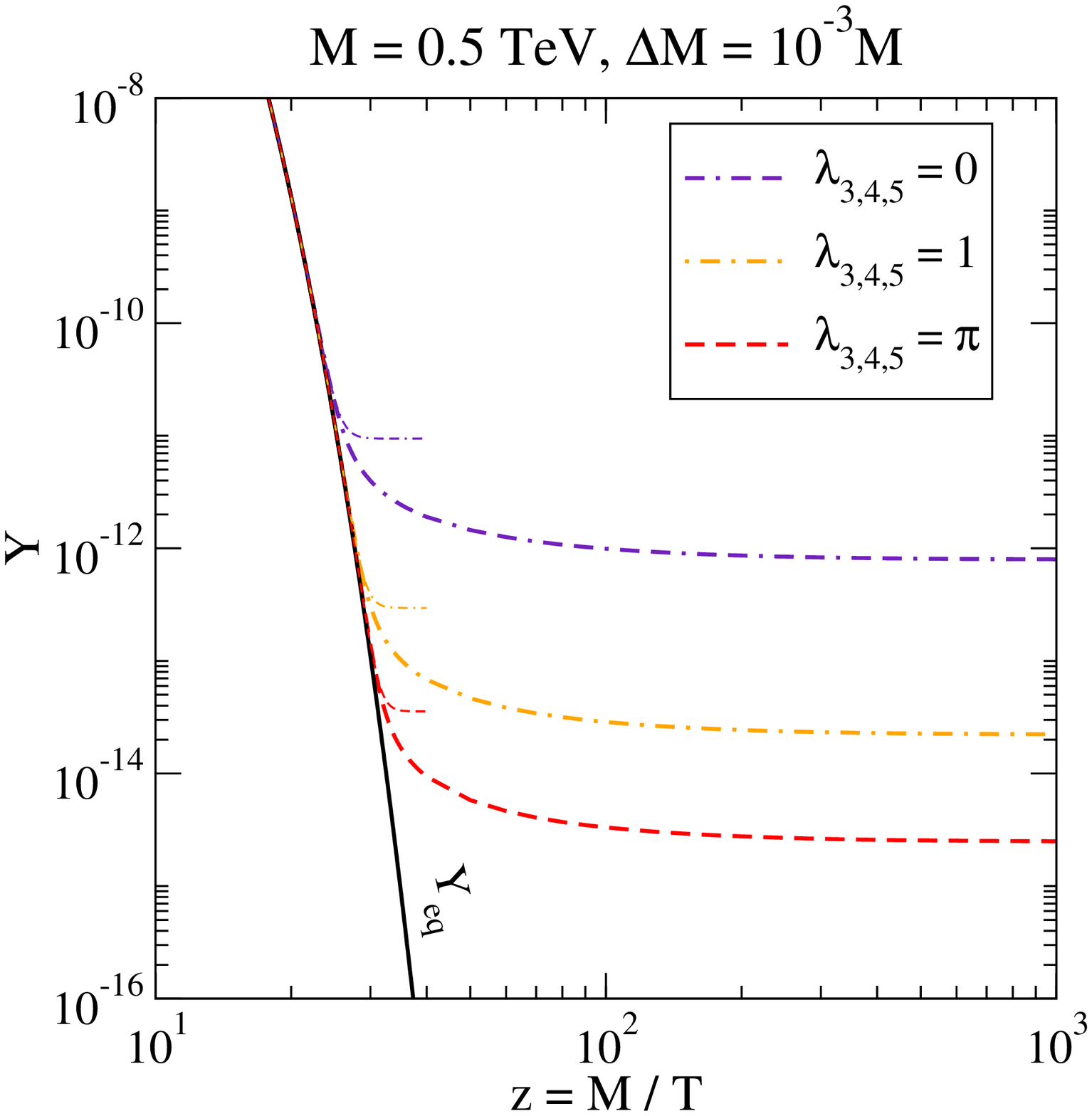}
 \hspace{0.1cm}
 \epsfxsize=5.0cm\epsfbox{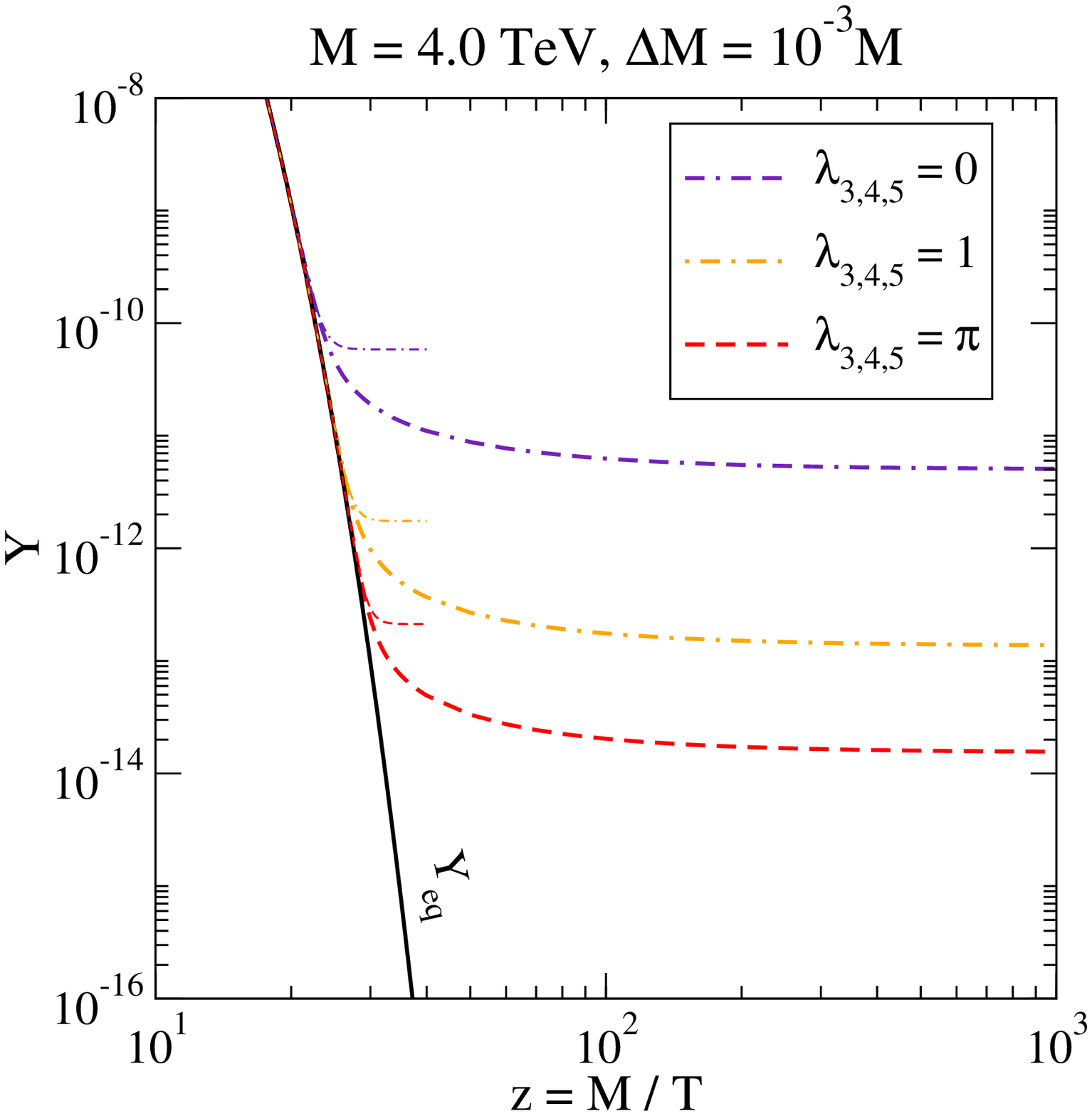}%
 \hspace{0.1cm}
 \epsfxsize=5.0cm\epsfbox{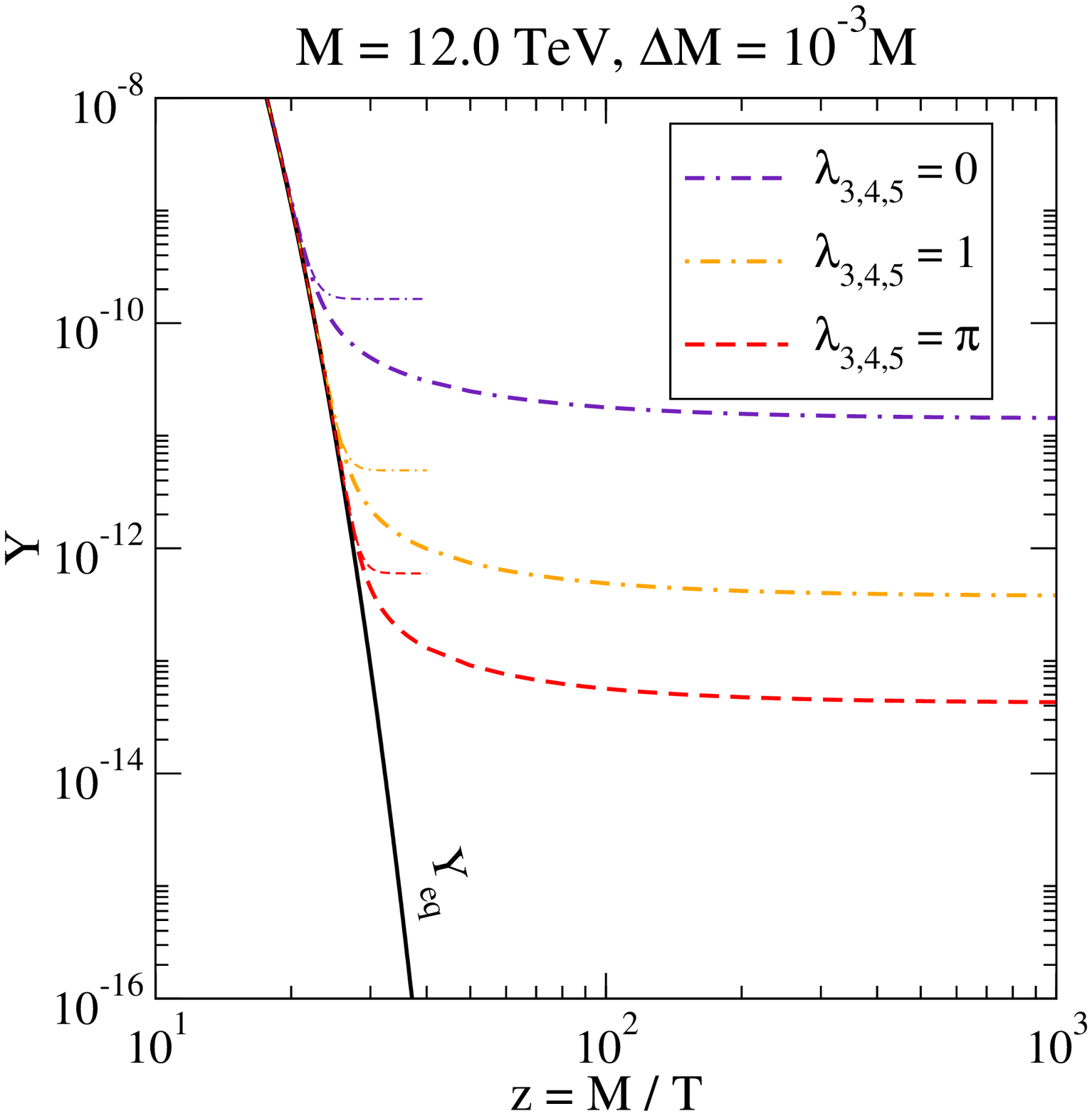}%
}

\caption[a]{\small
 The evolution of the yield parameter for various masses and  
 scalar couplings, as a function of $z\equiv M/T$. 
 The scalar couplings are evaluated at the scale $\bmu = 2M$, 
 we have set 
 $\lambda^{ }_3(2M) = \lambda^{ }_4(2M) = \lambda^{ }_5(2M)$, 
 and $\Delta M = 10^{-3} M$. 
 Thick lines correspond to the Lee-Weinberg equation in \eq\nr{boltzmann}, 
 and thin lines to the linearization in \eq\nr{response}, which is a good
 approximation for the initial decoupling. 
}

\la{fig:Y}
\end{figure}

As can be deduced from \fig\ref{fig:Y}, $Y^{ }_\rmi{eq}$ has become
exponentially small by the time that $z\sim 40$. 
In the absence of $Y^{ }_\rmi{eq}$, \eq\nr{dY} can be integrated into
\be
 \frac{1}{Y(z^{ }_\rmi{final})} - \frac{1}{Y(z=40)}
   = \int_{40}^{z^{ }_\rmi{final}}
   \frac{{\rm d}z}{z^2}
   \left.
   \frac{
   \langle \sigma^{ }_\rmi{eff}\, v \rangle M m^{ }_\rmi{Pl} \, 
   c(T)}{\sqrt{24\pi e(T)}} \right|^{ }_{T = M/z}
 \;. 
 \la{Yfinal}
\ee
The regime $z\gsim 40$ can easily 
reduce the dark matter abundance by a factor $2...3$. We choose 
$z^{ }_\rmi{final} = 10^4$ so that the contribution from late times
is typically at the percent level. Note that weak interactions are faster
than the Hubble rate 
down to $T\simeq 10$~MeV, so we may assume the dark matter
particles to be kinetically equilibrated in the whole $z$ range. 

It should however be noted that, taken literally,  
the growing Sommerfeld factor $\bar{S}^{ }_{1,\rmi{eff}}$ in 
\fig\ref{fig:barS} compromises the convergence of \eq\nr{Yfinal}
at large $z$. At the same time, 
at low temperatures kinetic and chemical equilibrium
is gradually lost in the dark sector,  
and the bound-state thermal abundance is presumably no longer 
available as an efficient annihilation channel once $\pi T \ll \alpha^2 M$. 
The value 
$z^{ }_\rmi{final} = 10^4$ represents a phenomenological compromise 
where the numerical effect from large $z$ is small, 
yet the physics assumptions
that went into the thermal analysis should still be 
intact. It would be interesting to 
understand the physics of this regime more precisely
(cf.\ also the comments in \ses\ref{ss:lowT} and \ref{se:concl}).

Eventually the heavier dark matter 
particles decay into the lightest one, so that
the final yield is $Y^{ }_\rmi{phys} = Y(z^{ }_\rmi{final})$.
The energy density carried by the lightest ones today is 
$\rho^{ }_\rmi{dm}(T^{ }_0) = M Y^{ }_\rmi{phys} s(T^{ }_0)$, 
and the energy fraction is
$
  \Omega^{ }_\rmi{dm}(T^{ }_0) =
        M Y^{ }_\rmi{phys} s(T^{ }_0)/\rho^{ }_\rmi{cr}(T^{ }_0)
$, 
where $\rho^{ }_\rmi{cr}$ is the 
current critical energy density. 
Inserting from ref.~\cite{pdg} 
$s(T^{ }_0) = 2\,891$/cm$^3$ and  
$ \rho^{ }_\rmi{cr}(T^{ }_0) = 1.0537 \times 10^{-5} h^2\, \mbox{GeV}$/cm$^3$  
yields 
\be
   \Omega^{ }_\rmi{dm} h^2 = \frac{M}{\mbox{GeV}}\,
      \frac{Y^{ }_\rmi{phys}}{3.645 \times 10^{-9}}
   \;, 
\ee 
which can be compared with the observed value
$ 
 \left. \Omega^{ }_\rmi{dm} h^2 \right|^{ }_\rmi{obs} = 
 0.1186(20)
$~\cite{planck}. 
Results are plotted in \fig\ref{fig:Omega}; 
a discussion is deferred to the first paragraph
of \se\ref{se:concl}.

\begin{figure}[t]
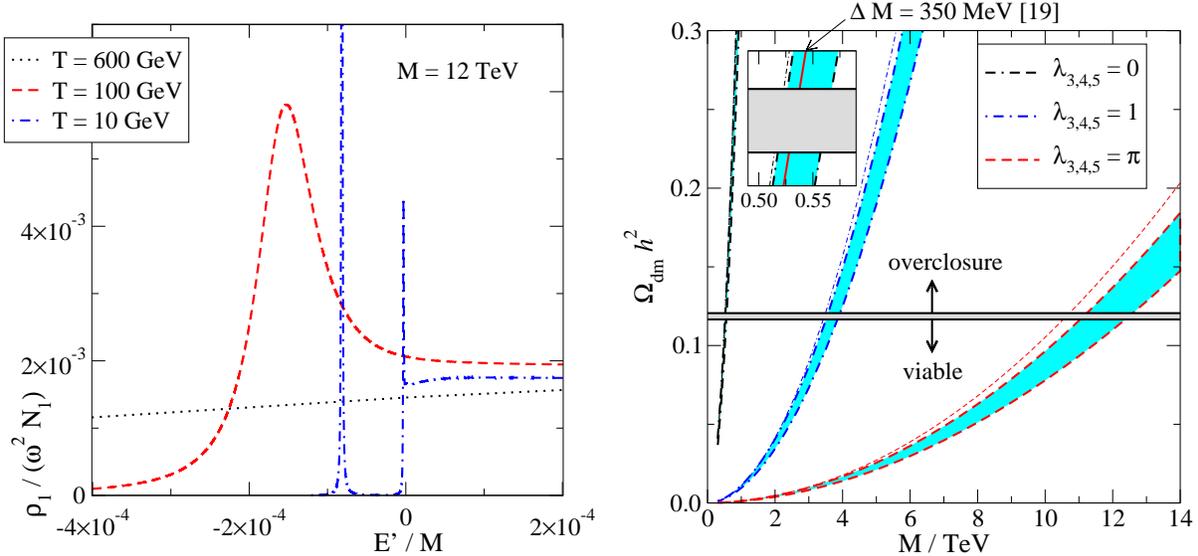


\hspace*{-0.1cm}
\centerline{%
 \epsfxsize=7.9cm\epsfbox{rho_M12000_V1.eps}
 \hspace{0.1cm}
 \epsfxsize=7.6cm\epsfbox{Omega.eps}
}

\caption[a]{\small
 Left:
 the spectral function $\rho^{ }_1$ for $M = 12$~TeV
 very close to threshold. A rapid broadening and merging of bound
 states can be observed as the temperature increases (the right-most
 peak is itself resolved into several peaks at lower temperatures). 
 Right: 
 the dark matter relic density, as a function of $M$/TeV, 
 for various quartic couplings. 
 Thin lines indicate the tree-level result; thick lines 
 the full result including thermal Sommerfeld and Salpeter corrections 
 and interaction rates. The error estimates of 
 \fig\ref{fig:barS} lead to modifications of the order of the
 thick line width, whereas the large uncertainties shown originate
 from varying the mass splitting in the range 
 $\Delta M = (0 ... 10^{-3})M$. 
 The horizontal line represents the observed value 
 $ 
  \left. \Omega^{ }_\rmi{dm} h^2 \right|^{ }_\rmi{obs} = 
  0.1186(20)
 $~\cite{planck}.  
}

\la{fig:Omega}
\end{figure}

%
\section{Conclusions and outlook}
\la{se:concl}

The purpose of this paper has been to illustrate and refine the general
formalism of ref.~\cite{threshold}, by applying it to a simple yet
phenomenologically viable dark matter computation. After the inclusion
of thermal effects, 
such as the Salpeter correction to
dark matter masses, the modification of the 
Sommerfeld effect through Debye screening, 
and thermal interaction rates, 
we find a conservative upper bound
for the mass of the lightest dark matter particle within the inert doublet
model (IDM), as a function of quartic scalar couplings. 
As a reference, we note 
that for vanishing quartic couplings  
values $M \lsim 535\pm 9$~GeV can typically be found in
literature
(cf.\ e.g.\ refs.~\cite{multiplet,ibarra}), 
and that for this case
we get $M \lsim 519\pm 4$~GeV by using free spectral functions
(cf.\ \fig\ref{fig:Omega}(right)). 
Switching on the thermally 
modified Sommerfeld factors, Salpeter corrections, and thermal
interaction rates, the bound increases to $M \lsim 523\pm 5$~GeV for
$\Delta M = 10^{-3} M$, and to  
$M \lsim 562\pm 5$~GeV for the extreme case $\Delta M/M \to 0$.   
For the maximal quartic couplings considered, 
$\lambda^{ }_3(2M) = \lambda^{ }_4(2M) = \lambda^{ }_5(2M) = \pi$, 
we obtain $M \lsim 10.6\pm 0.1$~TeV with free spectral functions; 
$M \lsim 11.1\pm 0.1$~TeV for $\Delta M = 10^{-3} M$; and 
$M \lsim 12.5\pm 0.1$~TeV for $\Delta M/M \to 0$. 
The uncertainties cited here originate from the observed
value of the dark matter relic density~\cite{planck}. 

In the high-mass regime 
the system displays a non-trivial 
bound-state spectrum at low temperatures (cf.\ \fig\ref{fig:Omega}(left)),
which leads to large Sommerfeld factors at large $z$
(cf.\ \fig\ref{fig:barS}). This results in efficient annihilation, 
and helps to push up the upper bound for $M$. We stress that 
the bound-state spectrum is easily addressed within our formalism, 
since the known Hard Thermal Loop resummed thermal interaction rate 
(reflecting the processes in \fig\ref{fig:processes}(right)) 
eliminates the need for complicated bound-state production and dissociation
rate computations. At very low temperatures, $T \ll \alpha^{3/2} M$, other
processes contribute as well (cf.\ \fig\ref{fig:processes}(left)), 
however these have also been studied in the QCD context 
(cf.\ refs.~\cite{jacopo2,nlo} and references therein), and the
same techniques could conceivably be generalized to cosmology. 
Once $\pi T \ll \alpha^2 M$, there is gradual departure from 
kinetic and chemical equilibrium in the dark sector, whose
study represents a complicated but interesting open problem.\footnote{%
 A nice recent investigation of non-equilibrium effects in another context 
 can be found in ref.~\cite{garny}. 
 }

Once the collider lower bound exceeds the 
cosmological upper bound
of \fig\ref{fig:Omega}(right), 
IDM is firmly excluded as a model, 
independently of astrophysical uncertainties related to
the local dark matter distribution. 
In practice, accepting modest astrophysical assumptions,  
direct and indirect non-detection constraints permit 
to set more stringent bounds than the overclosure one 
(cf.\ e.g.\ refs.~\cite{queiroz,ibarra} and references therein). 

One weakness of the IDM is that 
the quartic scalar couplings
can be varied in a broad range, which has a significant effect on 
the overclosure bound (cf.\ \fig\ref{fig:Omega}(right)). The quartic
couplings also influence mass splittings, resulting in 
a non-trivial multidimensional parameter dependence. 
If the couplings are large, their effects should
be resummed. For instance the scalar couplings 
affect the thermal corrections to dark matter masses; 
in contrast to the Salpeter correction
in \eq\nr{salpeter}, 
these effects are power-suppressed,  
$\Delta M^{ }_\rmii{$T$} \simeq 
(2\lambda^{ }_3+\lambda^{ }_4) T^2/(24 M)$. 
In addition, at $T \lsim 160$~GeV, the Higgs mechanism ($v > 0$)
generates cubic scalar couplings which lead to additional 
terms in the static potentials (cf.\ e.g.\ ref.~\cite{ibarra}). 
In the present investigation we resummed 
only effects from gauge couplings, which are not suppressed
by $T/M$ or $v/M$ and are therefore expected to
generically give the dominant contributions. 

Beyond the IDM, 
our interest lies in models including strongly 
interacting particles,  
which have attracted much recent interest in view
of the substantial role that bound states could play (cf.\ e.g.\ refs.~\cite{%
 threshold,4quark_lattice,%
 wimpo1,wimpo2,wimpo3,wimpo4,
 old3,old32,old35,old4,old45,old5,old52,old53,bound4,bound3,bound2,bound1,
 boundX,boundY
 }). 
Having now ``calibrated'' the formalism of ref.~\cite{threshold} through
a much-studied test case, 
we hope to address such models in the near future. 

%
\section*{Acknowledgements}

This work was supported by the Swiss National Science Foundation
(SNF) under grant 200020-168988. S.B.\ thanks 
Germano Nardini and Lewis Tunstall
for helpful discussions.

%
\appendix
\renewcommand{\thesection}{Appendix~\Alph{section}}
\renewcommand{\thesubsection}{\Alph{section}.\arabic{subsection}}
\renewcommand{\theequation}{\Alph{section}.\arabic{equation}}

%
\section{Explicit expressions for static potentials}

We present here the explicit expressions for the potentials appearing
in \eqs\nr{pot1}--\nr{pot3}. The potentials contain the Debye mass 
parameters defined in \eq\nr{Debye}, and the charged and neutral gauge
mass parameters~\cite{broken} 
\ba
 \mWt^2 & \equiv & \mW^2 + m_\rmii{E2}^2 \;, \\ 
 \mZt^2 & \equiv &  m_{+}^2 
  \;, \quad
  \mQt^2 \; \equiv \;  m_{-}^2  
  \;, \la{mZt_mQt} \\ 
  m_{\pm}^2 & \equiv & 
 \frac{1}{2} 
 \Bigl\{
   \mZ^2 + m_\rmii{E1}^2 + m_\rmii{E2}^2 \pm 
   \sqrt{\sin^2(2\theta) \mZ^4 +
   [\cos(2\theta) \mZ^2 + m_\rmii{E2}^2 - m_\rmii{E1}^2]^2}
 \Bigr\}
 \;. 
\ea 
The mixing angles are defined by
\ba
 \sin(2\theta^{ }_\rmii{}) 
 & \equiv & \frac{2g^{ }_1 g^{ }_2}{g_1^2 + g_2^2}
 \;, \\ 
 \sin(2\tilde\theta) 
 & \equiv &
 \frac{\sin(2\theta) \mZ^2 }{\sqrt{\sin^2(2\theta) \mZ^4 +
 [\cos(2\theta) \mZ^2 + m_\rmii{E2}^2 - m_\rmii{E1}^2]^2}}
 \;. \la{mixing} 
\ea
The neutral gauge field components are diagonalized as in 
\eqs(A.21)--(A.23) of ref.~\cite{threshold}. We define the functions
\ba
 \phi^{ }_r(m) & \equiv & 
 2 \int_0^\infty \! \frac{{\rm d}x }{(x^2+1)^2}
 \, \frac{\sin(x r m)}{r m} 
 \;, \la{phi_r} \\  
 \theta^{ }_r(m^{ }_1,m^{ }_2) & \equiv & 
 2 \int_0^\infty \! \frac{{\rm d}x }{x^2+1}
 \, \biggl[ \frac{\sin(x r m^{ }_1)}{r m^{ }_1}
 - \frac{\sin(x r m^{ }_2)}{r m^{ }_2}
 \biggr]
 \;. \la{theta_r}
\ea
Denoting furthermore $\tilde{c} \equiv \cos\tilde{\theta}$, 
$\tilde{s} \equiv \sin\tilde{\theta}$, 
$c \equiv \cos\theta$, and 
$s \equiv \sin\theta$, and renormalizing $r$-independent linear
divergences of the potentials as 
mentioned below \eq\nr{pot3}, we get
\ba
  \mathcal{V}^{ }_\rmii{${W}{W}$}(r)
 & = & 
 \frac{g_2^2}{16\pi}
 \biggl[
   \frac{\exp(-\mWt^{ }r)}{r}
  - \frac{i T  m_\rmii{E2}^2 \phi^{ }_r(\mWt^{ })}{\mWt^2}
 \biggr]
 \;, \la{VWWr} \\[2mm] 
  \mathcal{V}^{ }_\rmii{${W}{W}$}(0)
 & = & 
 - \frac{g_2^2}{16\pi}
 \biggl( \mWt^{ } + \frac{i T m_\rmii{E2}^2}{\mWt^2}  \biggr)
 + \left. \frac{g_2^2 \mW}{16\pi} \right|^{ }_{T = 0}
 \;, \la{VWW0} \\[2mm] 
  \mathcal{V}^{ }_\rmii{${A}{A}$}(r)  
 & = & 
 \frac{g_2^2}{16\pi}
 \biggl\{
 \frac{\tilde{s}^2 \exp(-\mQt^{ } r)}{r} +
 \frac{\tilde{c}^2 \exp(- \mZt^{ } r)}{r}
 -    
 i T \biggl[ 
  \frac{\tilde{s}^2(\tilde{c}^2 m_\rmii{E1}^2 + \tilde{s}^2 m_\rmii{E2}^2)
       \,\phi^{ }_r(\mQt^{ })}
       {\mQt^2}
 \nn 
 & + &  
  \frac{\tilde{c}^2(\tilde{s}^2 m_\rmii{E1}^2 + \tilde{c}^2 m_\rmii{E2}^2)
       \,\phi^{ }_r(\mZt^{ })}
       {\mZt^2}
 + \frac{2\tilde{c}^2\tilde{s}^2(m_\rmii{E2}^2 - m_\rmii{E1}^2)
       \,\theta^{ }_r(\mQt^{ },\mZt^{ })}
       {\mZt^2 - \mQt^2} 
 \biggr] \biggr\}
 \;, \\[2mm] 
  \mathcal{V}^{ }_\rmii{${A}{A}$}(0)  
 & = & 
 -\frac{g_2^2}{16\pi}
 \biggl\{
 \tilde{s}^2 \mQt^{ }+ \tilde{c}^2 \mZt^{ }
 +    
 i T \biggl[ 
  \frac{\tilde{s}^2(\tilde{c}^2 m_\rmii{E1}^2 + \tilde{s}^2 m_\rmii{E2}^2)}
       {\mQt^2}
 \nn 
 & + &  
  \frac{\tilde{c}^2(\tilde{s}^2 m_\rmii{E1}^2 + \tilde{c}^2 m_\rmii{E2}^2)}
       {\mZt^2}
 + \frac{2\tilde{c}^2\tilde{s}^2(m_\rmii{E2}^2 - m_\rmii{E1}^2)}
       {\mZt^2 - \mQt^2} \ln\biggl( \frac{\mZt^2}{\mQt^2} \biggr)
 \biggr] \biggr\}
 + \left. \frac{g_2^2 c^2 \mZ}{16\pi} \right|^{ }_{T = 0}
 \;, \hspace*{1cm} \la{VAA0} \\[2mm] 
  \mathcal{V}^{ }_\rmii{${B}{B}$}(r) 
 & = & 
 \frac{g_1^2}{16\pi}
 \biggl\{
 \frac{\tilde{c}^2 \exp(-\mQt^{ } r)}{r} +
 \frac{\tilde{s}^2 \exp(- \mZt^{ } r)}{r}
 -    
 i T \biggl[ 
  \frac{\tilde{c}^2(\tilde{c}^2 m_\rmii{E1}^2 + \tilde{s}^2 m_\rmii{E2}^2)
       \,\phi^{ }_r(\mQt^{ })}
       {\mQt^2}
 \nn 
 & + &  
  \frac{\tilde{s}^2(\tilde{s}^2 m_\rmii{E1}^2 + \tilde{c}^2 m_\rmii{E2}^2)
       \,\phi^{ }_r(\mZt^{ })}
       {\mZt^2}
 + \frac{2\tilde{c}^2\tilde{s}^2(m_\rmii{E1}^2 - m_\rmii{E2}^2)
       \,\theta^{ }_r(\mQt^{ },\mZt^{ })}
       {\mZt^2 - \mQt^2} 
 \biggr] \biggr\}
 \;, \hspace*{1cm} \\[2mm] 
  \mathcal{V}^{ }_\rmii{${B}{B}$}(0) 
 & = & 
 -\frac{g_1^2}{16\pi}
 \biggl\{
 \tilde{c}^2 \mQt^{ }+ \tilde{s}^2 \mZt^{ }
 +    
 i T \biggl[ 
  \frac{\tilde{c}^2(\tilde{c}^2 m_\rmii{E1}^2 + \tilde{s}^2 m_\rmii{E2}^2)}
       {\mQt^2}
 \nn 
 & + &  
  \frac{\tilde{s}^2(\tilde{s}^2 m_\rmii{E1}^2 + \tilde{c}^2 m_\rmii{E2}^2)}
       {\mZt^2}
 + \frac{2\tilde{c}^2\tilde{s}^2(m_\rmii{E1}^2 - m_\rmii{E2}^2)}
       {\mZt^2 - \mQt^2} \ln\biggl( \frac{\mZt^2}{\mQt^2} \biggr)
 \biggr] \biggr\}
 + \left. \frac{g_1^2 s^2 \mZ}{16\pi} \right|^{ }_{T = 0}
 \;.  \la{VBB0}
\ea

The potentials get considerably simplified in the short-distance
limit $r \ll 1/ \mZt^{ }$. 
Then their divergent $r$-dependent parts read
\be
 \mathcal{V}^{ }_\rmii{${W}{W}$}(r) \; \simeq \; \frac{g_2^2}{16\pi r}
 \;, \quad
 \mathcal{V}^{ }_\rmii{${A}{A}$}(r) \; \simeq \; \frac{g_2^2}{16\pi r}
 \;, \quad
 \mathcal{V}^{ }_\rmii{${B}{B}$}(r) \; \simeq \; \frac{g_1^2}{16\pi r}
 \;,
\ee
and \eqs\nr{pot1}--\nr{pot3} become
\be
 \mathcal{V}^{ }_1(r)  \; \simeq \; - \frac{3 g_2^2 + g_1^2}{16\pi r}
 \;, \quad
 \mathcal{V}^{ }_2(r)  \; \simeq \; \frac{g_2^2 - g_1^2}{16\pi r}
 \;, \quad
 \mathcal{V}^{ }_{3,4}(r)  \; \simeq \; \frac{g_2^2 + g_1^2}{16\pi r}
 \;.
\ee
Defining 
$
 \alpha^{ }_1 \equiv 
 (3 g_2^2 + g_1^2)/(16\pi)
$, 
$
 \alpha^{ }_2 \equiv 
 (g_2^2 - g_1^2)/(16\pi)
$
 and 
$
 \alpha^{ }_{3,4} \equiv 
 (g_2^2 + g_1^2)/(16\pi)
$, 
the corresponding Sommerfeld factors read~\cite{fadin}
\be
  S^{ }_1 = \frac{ X^{ }_1 } { 1 - e ^{ - X^{ }_1 } }
  \;, \quad 
  S^{ }_{2,3,4} = \frac{ X^{ }_{2,3,4} } { e ^{ X^{ }_{2,3,4} } - 1 }
  \;, \la{S}
\ee
where $X^{ }_i \equiv \pi \alpha^{ }_i /v$ and $v$ parametrizes 
$E'$ from \eq\nr{Laplace1} as $E' = 2 \Delta M^{ }_\rmii{$T$} + M v^2$.

We note that \eqs\nr{VWWr}--\nr{VBB0} are based on 
evaluating gauge field self-energies in the Hard Thermal Loop
approximation. This is justified as long as the particles with 
which gauge fields interact are ultrarelativistic, 
i.e.\ with masses $m \ll \pi T$. 
If $m \gsim \pi T$, the self-energies take a more complicated
form (cf.\ appendix~A of ref.~\cite{threshold} for the full 1-loop 
self-energy
matrix of the neutral components $A^3_0,B^{ }_0$), and thermal 
modifications cannot be captured by the two Debye mass
parameters $m_\rmii{E1}^2$ and  $m_\rmii{E2}^2$. Nevertheless,
it is possible to identify the light-fermion contribution to 
the Debye masses. If we consider vanishing spatial momentum; 
model top and bottom quarks by a common ``fermionic'' mass $m^{ }_\ff$; 
and model $W^\pm$, $Z^0$ and Higgs bosons by a common
``gauge'' mass $m^{ }_g$; 
then \eq(A.6) of ref.~\cite{broken} 
shows that terms mixing $A^3_0$ and $B^{ }_0$
drop out, and we may replace \eq\nr{Debye} with
\be
  m^{2}_\rmii{E1} \; \simeq \; 
 \frac{g_1^2}{2} 
 \biggl[ \frac{49T^2}{18}
  + \frac{11 \chi^{ }_\rmii{F}(m^{ }_{\ff})}{3}
  + \chi^{ }_\rmii{B}(m^{ }_g)
 \biggr]  
 \;, \quad
 m^{2}_\rmii{E2} \; \simeq \; 
 \frac{g_2^2}{2} 
 \biggl[ \frac{3T^2}{2}
  + 3 \chi^{ }_\rmii{F}(m^{ }_\ff)
  + 5\chi^{ }_\rmii{B}(m^{ }_g)
 \biggr] 
 \;. \la{Debye_model}
\ee
Here the fermionic and bosonic susceptibilities read 
\be
 \chi^{ }_\rmii{F}(m^{ }_\ff) \;\equiv\; 
 \int_\vec{p} \bigl[ - 2 \nF{}'(E^{ }_\ff) \bigr] 
 \;\; \stackrel{m^{ }_\ff\to 0}{\to} \;\; \frac{T^2}{6}
 \;, \quad
 \chi^{ }_\rmii{B}(m^{ }_g) \;\equiv\; 
 \int_\vec{p} \bigl[ - 2 \nB{}'(E^{ }_g) \bigr] 
 \;\; \stackrel{m^{ }_g\to 0}{\to} \;\; \frac{T^2}{3}
 \;, \la{suscs}
\ee
where $\nF{}$ and $\nB{}$ are the Fermi and Bose distributions, 
respectively. 
We have adopted \eq\nr{Debye_model} for modelling 
the low-temperature regime, inserting
$m^{ }_\ff \simeq (m^{ }_t m^{ }_b)^{1/2}$ and 
$m^{ }_g \simeq (\mZ^{ } \mW^2 m_\phi)^{1/4}$, 
but stress that this represents a purely 
phenomenological recipe within the complicated 
temperature interval $m^{ }_b \lsim \pi T \lsim m^{ }_t$. 

\small{
%

}

\end{document}